\documentclass[fleqn,usenatbib]{mnras}

\usepackage{newtxtext,newtxmath}
\usepackage{graphicx}
\usepackage{amsmath} 
\usepackage{gensymb}
\usepackage{enumerate}
\usepackage{anyfontsize}

\usepackage[T1]{fontenc}
\DeclareRobustCommand{\VAN}[3]{#2}
\let\VANthebibliography\thebibliography
\def\thebibliography{\DeclareRobustCommand{\VAN}[3]{##3}\VANthebibliography}

\title[Measuring substructure power spectrum with CNNs]{Measuring the Substructure Mass Power Spectrum of 23 SLACS Strong Galaxy-Galaxy Lenses with Convolutional Neural Networks}

\author[J. Fagin et al.]{
Joshua Fagin$^{1,2,3}$\thanks{jfagin@gradcenter.cuny.edu},
Georgios Vernardos$^{2,3}$,
Grigorios Tsagkatakis$^{4}$,
Yannis Pantazis$^{5}$,
Anowar J. Shajib$^{6,7,8}$,
\newauthor
Matthew O'Dowd$^{1,2,3}$
\\
$^{1}$The Graduate Center of the City University of New York, 365 Fifth Avenue, New York, NY 10016, USA\\
$^{2}$Department of Astrophysics, American Museum of Natural History, Central Park West and 79th Street, NY 10024-5192, USA\\
$^{3}$Department of Physics and Astronomy, Lehman College of the CUNY, Bronx, NY 10468, USA\\
$^{4}$Institute of Computer Science, FORTH, GR-70013, Heraklion, Greece\\
$^{5}$Institute of Applied and Computational Mathematics, FORTH, GR-70013, Heraklion, Greece\\
$^{6}$Department of Astronomy \& Astrophysics, University of Chicago, Chicago, IL 606374, USA \\
$^{7}$Kavli Institute for Cosmological Physics, University of Chicago, Chicago, IL 60637, USA \\
$^{8}$NHFP Einstein Fellow \\
}

\date{Accepted XXX. Received YYY; in original form ZZZ}

\pubyear{2023}

\begin{document}
\label{firstpage}
\pagerange{\pageref{firstpage}--\pageref{lastpage}}
\maketitle

\begin{abstract}
Strong gravitational lensing can be used as a tool for constraining the substructure in the mass distribution of galaxies. In this study we investigate the power spectrum of dark matter perturbations in a population of 23 \emph{Hubble Space Telescope} images of strong galaxy-galaxy lenses selected from The Sloan Lens ACS (SLACS) survey. We model the dark matter substructure as a Gaussian Random Field perturbation on a smooth lens mass potential, characterized by power-law statistics. We expand upon the previously developed machine learning framework to predict the power-law statistics by using a convolutional neural network (CNN) that accounts for both epistemic and aleatoric uncertainties. For the training sets, we use the smooth lens mass potentials and reconstructed source galaxies that have been previously modelled through traditional fits of analytical and shapelet profiles as a starting point. We train three CNNs with different training set: the first using standard data augmentation on the best-fitting reconstructed sources, the second using different reconstructed sources spaced throughout the posterior distribution, and the third using a combination of the two data sets. We apply the trained CNNs to the SLACS data and find agreement in their predictions. Our results suggest a significant substructure perturbation favoring a high frequency power spectrum across our lens population.
\end{abstract}


\section{Introduction}

Dark matter is a currently unknown component of the Universe that does not interact electromagnetically with baryonic matter but is still influenced by gravitational fields. The theory of Cold Dark Matter (CDM) and dark energy have successfully explained the large-scale structure and expansion of the Universe in both observed~\citep{Cosmology1,Cosmology2} and simulated~\citep{Cosmology3,Cosmology4} data. On the small cosmological scale of $<1$ Mpc, however, different dark matter models diverge~\citep{Dark_matter_on_small_scale} challenging the CDM paradigm. 

Strong gravitational lensing occurs when light from a distant source galaxy is deflected by a closer galaxy along the observer's line of sight forming arcs, rings, and multiple images~\citep[an overview on the theory and goals of strong gravitational lensing physics can be found in][]{shajib2022strong}. Dark matter can be studied by analysing its effect on the shape and smoothness of the gravitational lens potential of the lensing galaxy~\citep{Koopmans_2005,Vegetti2009}. It has been found that massive elliptical galaxies are very close to being isothermal~\citep{Isothermal1,Isothermal2,Isothermal3,Isothermal4}, but the presence of compact dark substructures of order $10^8 \text{M}_\odot$ has been detected~\citep[e.g.][]{Vegetti_2010,Subhalo_mass2,Subhalo_mass3}.

Currently several hundred strong galaxy-galaxy lensed systems have been observed, with the most complete survey being the Sloan Lens ACS (SLACS) Survey~\citep{Sloan_ACS_lenses} that includes 85 lenses. This number is bound to greatly increase in the near future with the European Space Agency’s \textit{Euclid} telescope~\citep{Euclid} that launched July 2023 and the Vera C. Rubin Observatory~\citep{LSST} expected to be operational in 2024. Together they are projected to observe billions of galaxies including tens of thousands of strongly lensed systems~\citep{Number_Strong_Lens_Discovered}. 

Traditional lens modelling relies on using parametric models, where the lens and source parameters are optimized via Markov Chain Monte Carlo (MCMC) sampling~\citep{MCMC_lens_modeling}. Various techniques have emerged such as reconstructing the source on a grid of pixels~\citep{Semilinear_inversion,Vegetti2009,Nightingale_2021,galan2021,Vernardos2022} or with a basis set of shapelets~\citep{Shapelets,Lens_modeling_with_basis,Shapelets_for_lense_modeling}. These techniques are computationally demanding, as they require sampling across a large non-linear parameter space for each lens. Modelling one lens can take weeks or months of computational time and often requires additional expert intervention~\citep{Schuldt2021}. Hence applying MCMC to the tens of thousands of lenses anticipated in upcoming surveys would prove challenging. With such a large number of strongly lensed systems expected in the near future, there is a need for fast and automated methods to analyse them all.

Machine learning (ML) methods excel at efficiently analysing large quantities of data and have been successfully applied to a number of problems in strong gravitational lensing. Convolutional neural networks (CNNs) are particularly well suited for image data and are now the standard method for strong gravitational lens detection \citep{Strong_Lensing_Detection_With_CNN_2,Strong_Lensing_Detection_With_CNN_1,Strong_Lensing_Detection_With_CNN_4,Strong_Lensing_Detection_With_CNN_3,CNN_lens_detection,Rojas_2022,Savary2022}. A fully trained neural network could model lenses in a fraction of a second per system, enabling easy application to tens of thousands of lenses. 
\citet{CNN_without_uncertainties} first demonstrated how a CNN could be used to estimate the strong lens mass model parameters and introduced a method for obtaining uncertainties on those predictions in~\citet{Uncertainty_in_SIE_predictions}. Since then there have been several other examples of lens modelling using CNNs~\citep{Estimating_SIE_Par_Using_CNN,Schuldt2021,Peason2021,CNN_Hierarchical_Inference}. CNNs have also been applied to strong lensing problems involving dark matter subhaloes~\citep{Dark_matter_CNN_1,Dark_matter_CNN_2,Dark_matter_CNN_3,Subhalo_mass_using_machine_learning,Domain_adaptation,CNN_Substructure_Mass_Function,CNN_Substructure_Mass_Function2,Adam_2023}.
The method proposed by~\citet{Biggio2022} for example, is a new approach that combines traditional sampling and parametric techniques with ML, a fully-connected network in this case, that is used to model the unknown prior on the source or small-scale lens potential.

A common assumption in dark matter studies with lensing is treating its effect as a perturbation on a smooth lens potential.
Although massive, isolated subhaloes can be modelled individually and have a localized effect, the much more common smaller mass subhaloes permeate the field of view and have a collective effect on the lensed features of the source.
This has led many authors to treat such perturbations as a Gaussian Random Field~\citep[GRF;][]{Chatterjee_2017,Vernardos2022,Galan2022,Biggio2022, Bayer_2023_1, Bayer_2023_2}.
Alternatively one could use a parametric approach where dark matter substructure is modelled as an ensemble of individual subhaloes with Navarro–Frenk–White~\citep[NFW;][]{NFW} profiles~\citep[e.g.,][]{CNN_Substructure_Mass_Function,CNN_Substructure_Mass_Function2}. In this work we adopt the statistical approach of treating the lens perturbation as a GRF parameterized by its power-law statistics.

\citet{Chatterjee_2017} first proposed a method for measuring the GRF power-law in simulated data by analysing the surface brightness anomalies in the lens, which are the residuals between the smooth lens model reconstruction and the data. They then predict the GRF power-law parameters through an MCMC analysis of the power spectrum. \citet{Bayer_2023_1} further developed the method of measuring the power spectrum of surface brightness anomalies and applied it to a sample of ten SLACS lenses, carefully modelling the noise properties of each lens to predict the power spectrum. As a pilot application, \citet{Bayer_2023_2} were able to obtain upper bound constraints on the GRF power spectrum for the single SLACS lens SDSS J0252+0039, selected due to its simple geometry and high signal-to-noise ratio. Recently, \citet{Vernardos2022,Galan2022,Biggio2022} also showed how a GRF perturbation to the lens potential could be recovered in mock data using three different techniques, but they have not yet been applied to real observations.

A common ML method of obtaining uncertainties on parameter predictions is formulated in~\citet{kendall2017uncertainties} and applied to strong lensing modelling in~\citet{Uncertainty_in_SIE_predictions} that gives an approximate Bayesian framework for strong lens modelling with uncertainties using variational inference. The method involves minimizing a negative Gaussian log-likelihood loss function to produce the mean and variance for each prediction and uses Monte Carlo dropout as a method to incorporate the uncertainty in the model weights. A different approach was employed by~\citet{Vernardos_2020}, who introduced a novel uncertainty-aware neural network framework to quantify the uncertainty in dark matter power spectrum parameter predictions. Their method involves training a CNN to produce discrete probability vectors in the parameter space that can also quantify the uncertainty in the predictions. The CNN is trained by using uniform probability distributions as target distributions (the training labels) instead of using the true values directly. The CNN never knows the ground truth values, since they can be anywhere within the target distribution. The main advantage of this ML framework is that there can be arbitrary uncertainty in the input labels and the output is a non-parametric probability distribution.

In this work, we expand upon the uncertainty framework introduced in~\citet{Vernardos_2020} and apply it to a sample of 23 observed strong galaxy-galaxy lenses selected from the Sloan Lens ACS (SLACS) Survey~\citep{Sloan_ACS_lenses}. \citet{Vernardos_2020} used only three distinct source galaxies to generate a mock data set to train their CNN, and it was found that the results were not generalizable to other sources. \citet{Shajib_2021} conducted traditional MCMC lens modelling on the same SLACS sample and provided reconstructed source brightness profiles. By using these reconstructed sources to build our training set, we ensure that our CNN is trained with mock lenses closely resembling the real data. We create three mock data sets to train separate CNNs: the first applies standard data augmentation techniques to the best-fitting sources, the second employs different sources from various stages of the MCMC chain, and the third combines the previous two data sets. We then apply the pretrained CNNs to the SLACS lenses and average the resulting predictions to better generalize our ultimate prediction to the SLACS data.

We first present our selected sample of SLACS lenses in Section~\ref{lens_sample_sec}. We then describe the process of generating simulated strongly lensed images in Section~\ref{method}, followed by the training of an uncertainty-aware CNN to predict the substructure mass power spectrum parameters in Section~\ref{CNN_section}. We apply the trained CNNs to our SLACS lens sample and compare the results in Sect~\ref{Application_to_real_data_section}. We offer our concluding remarks and discussion in Section~\ref{Conclusion_sec}.

\begin{figure*}
    \centering
    \includegraphics[width=0.97\textwidth]{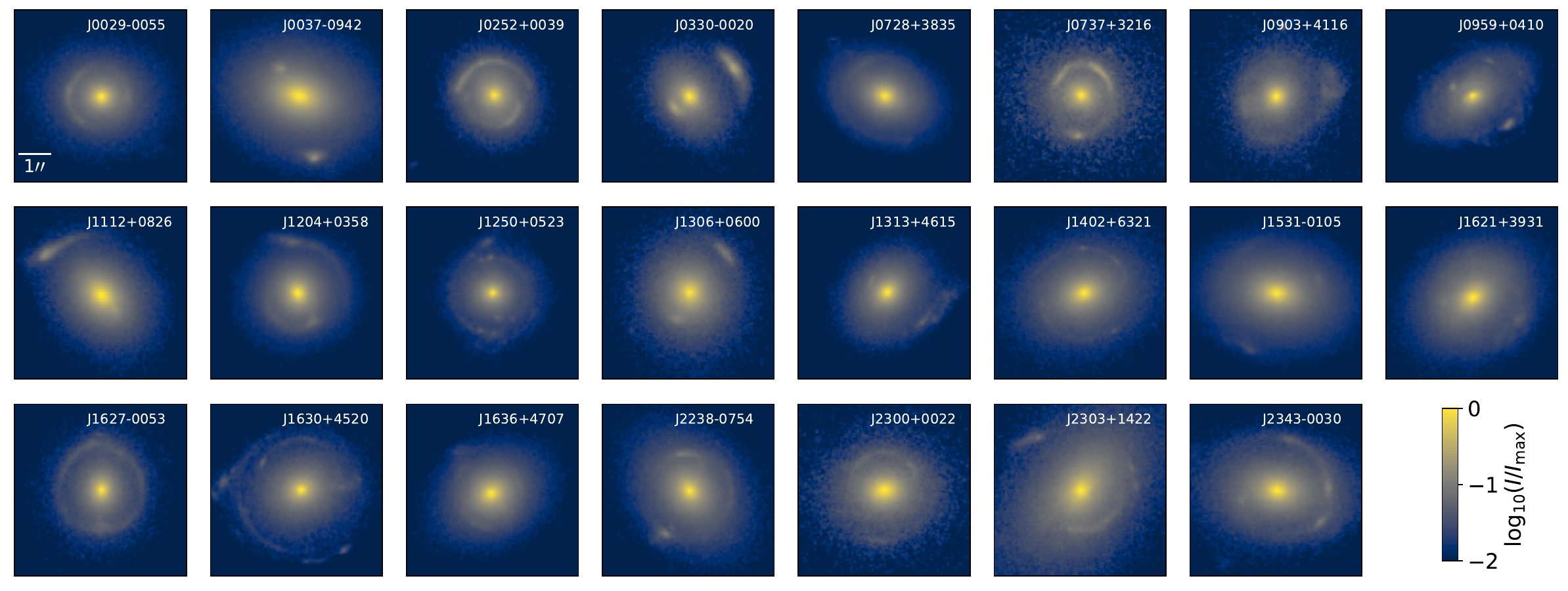}
    \caption{Our lens sample of 23 \emph{HST} images of SLACS lenses~\citep{Sloan_ACS_lenses} taken in the visible bands F555W or 5606W. Each image is scaled by its maximum brightness to better distinguish the arcs and rings of the lensed source from the lens light. All panels have the same dimensions, i.e. $5.5''$ on a side with a resolution of $0.05''$ and $1''$ shown in the top leftmost panel.}
    \label{real_data}
\end{figure*}

\section{SLACS lens sample}
\label{lens_sample_sec}

Our lens sample consists of 23 \emph{Hubble Space Telescope} (\emph{HST}) images obtained by the Sloan Lens ACS Survey~\citep{Sloan_ACS_lenses}, the same sample used by~\citet{Shajib_2021}. Their original sample consisted of 50 out of the 85 SLACS lenses selected on the basis of having simpler source shapes, not having nearby satellite galaxies along the line of sight, and not being disc-like (see section 2 of their paper for more details). They were able to successfully model 23 out of the 50 selected lenses.

The selected lenses were observed by either the Advanced Camera for Surveys (ACS) or the Wide Field and Planetary Camera 2 (WFPC2). The WFPC2 images were reduced by the original SLACS analysis~\citep{Sloan_ACS_lenses}, and~\citet{Shajib_2021} reduced the ACS images using the \texttt{AstroDrizzle}\footnote{\href{https://github.com/spacetelescope/drizzlepac}{https://github.com/spacetelescope/drizzlepac}} software package~\citep{Astro_drizzel_package}. Each image has a width of $5.5''$ with a resolution of $0.05''$ per pixel. Figure~\ref{real_data} shows the 23 lensed images used in this work.

\section{Lens simulation}
\label{method}

In the absence of a large set of real observations, as is the case in strong lensing, a synthetic training set should be designed to be broad enough to include a large range of possible lenses but also include images that are very close to the observations.
\citet{Vernardos_2020} showed that the most important factor in this is the source brightness profile. We address this by using the reconstructed sources obtained by~\citet{Shajib_2021} to create training sets that are representative of our SLACS lens sample. In this section we present how we generate our training sets using these sources, as well as our assumptions on the lens mass, lens light, and instrumental effects.

\subsection{Smooth lens potential}
\label{Smooth_lens_sec}

Our method can be used for any smooth lens model, but here we select a Singular Isothermal Ellipsoid~\citep[SIE;][]{SIE_1993,SIE_1994,SIE_1998} as it has been shown that massive elliptical galaxies, like those in the SLACS sample, are close to isothermal~\citep{Isothermal1,Isothermal2,Isothermal3,Isothermal4}. The SIE convergence is defined by:
\begin{equation}
\kappa(x,y) = \frac{b}{2\sqrt{q^2{x\prime}^2+{y\prime}^2}} ,
\end{equation}
where $b$ is the Einstein radius of the lens related to the strength of the lens potential and $q$ is the minor to major axis ratio. The coordinates ($x\prime$,$y\prime$) include a rotation by a position angle $\theta_{\mathrm{pa}}$ and transverse shift by $(x_0,y_0)$. We also include an external shear that is parameterized by its magnitude $\gamma$ and direction $\phi$. Our smooth lens model is the sum of the SIE and external shear deflection potentials and has seven free parameters. The best-fitting values for these parameters have been measured in~\citet[][table 1]{Shajib_2021}.
The parameter space in our simulation is chosen to encompass the full range of all the selected SLACS lenses and is given in Table~\ref{Parameter_table}.

\begin{table}
\centering
\caption{Lens mass, lens light, and noise parameters and ranges for generating the mock data set. Each parameter is drawn uniformly within the given range. The lens light is modelled as an elliptical Sérsic profile with mean location, orientation, and axis ratio correlated to the lens mass model parameters $\{x_0,y_0,q,\theta_{\mathrm{pa}}\}$ but deviated by a Gaussian random variable with a standard deviation of $2.5$ per cent the maximum range (see Section~\ref{lens_galaxy_light_sec}). The effective radius $r_{\mathrm{eff}}$ of the Sérsic profile is proportional to the chosen Einstein radius $b$. The pixelated sources are described in Section~\ref{Source_galaxy_profiles_section}. We also randomly select a point spread function to convolve with our simulated image from the 23 available (see Section~\ref{noise_sec}).}
 \begin{tabular}{l l c c} 
Name & Description & Min. & Max. \\
 \hline\hline
 \multicolumn{4}{l}{Lens model parameters}\\
 \hline
$b$ & Einstein radius & $0.8''$ & $1.8''$\\
$q$ & axis ratio & $0.5$ & $1.0$\\
$\theta_{\mathrm{pa}}$ & position angle & $0.0\degree$ & $180\degree$ \\
$x_0$ & lens centre & $-0.1''$ & $0.1''$ \\ 
$y_0$ & lens centre & $-0.1''$ & $0.1''$ \\
$\gamma$ & shear magnitude & 0.0 & 0.12 \\ 
$\phi$ & shear direction & $0.0\degree$ & $180\degree$ \\
$\log_{10}(\sigma_{\mathrm{\delta\psi}}^2)$ & GRF strength & $-5.0$ & $-2.0$ \\
$\beta$ & GRF power-law & $3.0$ & $8.0$ \\
\hline
\multicolumn{4}{l}{Additional lens light parameters}\\
\hline
$n$ & Sérsic index & $3.0$ & $8.0$  \\
$r_{\mathrm{eff}}$ & half-light radius & $0.5b$ & $2.6b$ \\
$\log_{10}(R)$ & lens light ratio & $-0.1$ & $2.0$ \\
 \hline
\multicolumn{4}{l}{Noise parameters}\\
 \hline
$L_{\mathrm{cor}}$ & noise correlation & $0.0''$ & $0.05''$ \\
$\log_{10}(\sigma_{\mathrm{noise}})$ & noise level & $-3.25$ & $-2.25$\\
\end{tabular}
\label{Parameter_table}
\end{table}

\begin{figure*}
\centering
\includegraphics[width=0.97\textwidth]{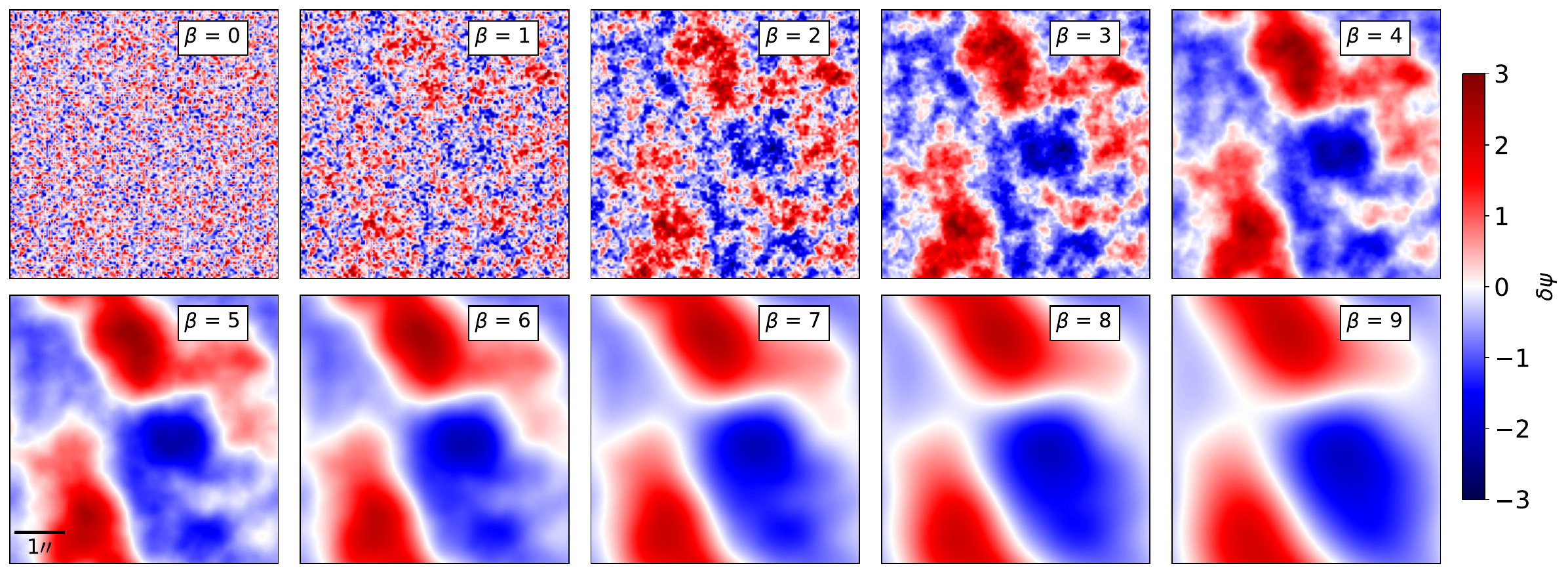}
\caption{Realizations of a GRF with the same random noise realization but different power spectrum slopes $\beta$ and fixed variance of the fluctuations $\sigma_{\delta\psi}^2 = 1$. All panels have the same dimensions, i.e. $5.5''$ on a side with a resolution of $0.05''$ and $1''$ shown in the bottom leftmost panel.} 
\label{GRF_vary_beta}
\end{figure*}

\subsection{GRF perturbations}
\label{Perturbed_lens_sec}

We model the substructure perturbation on the lens potential as a GRF, denoted $\delta\psi$. The GRF is a stochastic field with characteristic power spectrum parameterized by a power-law:
\begin{equation} \label{power_law}
P(k) = A k^{-\beta} ,
\end{equation}
where $A$ is the amplitude that scales with the variance of the zero mean $\delta\psi$, $\beta$ is the power-law coefficient, and $k$ is the wavenumber of the Fourier harmonic. The amplitude $A$ is related to the variance of the GRF by:
\begin{equation} \label{power_law_amp}
A(\sigma_{\delta\psi},\beta,L) = \frac{L^2\sigma_{\delta\psi}^2}{\sum_{k_x}\sum_{k_y} k^{-\beta}} ,
\end{equation}
where $L$ is the length of each side of the image, $\sigma_{\delta\psi}^2$ is the variance of the fluctuation, and \mbox{$k = \sqrt{k_x^2+k_y^2}$} is the magnitude of the Fourier wavenumber. We follow the convention of~\citet{Bayer_2023_2} and predict the log-variance of the GRF $\log_{10}(\sigma_{\delta\psi}^2)$ as a proxy for the amplitude of the power-law $A$, since it is independent of the scale of the image $L$ and the power-law slope $\beta$. Equation~(\ref{power_law_amp}) can later be used to transform our predicted $\log_{10}(\sigma_{\delta\psi}^2)$ back to $A$. 

Figure~\ref{GRF_vary_beta} shows instances of a GRF for different values of $\beta$. It can be seen that the power-law slope $\beta$ determines the smoothness or graininess of the GRF perturbation. The residuals between a perturbed and unperturbed lens potential increases with larger $\sigma_{\delta\psi}$ and with smaller $\beta$. In~\citet[][fig. 5]{Vernardos_2020}, they show how the residuals between a perturbed and unperturbed mock lens vary as a function of $A$ and $\beta$. When $\beta = 0$, the GRF becomes normal Gaussian white noise where each pixel is uncorrelated.

We use the range of $\beta\in [3,8]$ consistent with~\citet{Bayer_2023_2} and~\citet{Vernardos_2020}. This is motivated by the observation that the GRF does not change significantly when $\beta > 8$, as shown in Fig.~\ref{GRF_vary_beta}. At $\beta < 3$ the GRF appears more granular than expected for dark matter subhaloes, resembling random, slightly correlated noise. We take the range of $\log_{10}(\sigma_{\delta\psi}^2) \in [-5,-2]$ based on the results of~\citet[][fig. 9]{Bayer_2023_2} where they showed the exclusionary probability for lens SDSS J0252+0039 within that range. This is also approximately the same scale as~\citet{Vernardos_2020} where they considered $\log_{10}(A) \in [-5,-2]$.

We calculate the GRF on a grid of pixels with the same size and resolution as the SLACS observations, and the gradient is calculated numerically to produce the deflection angles.
Our total lens model is the sum of the smooth lens model and the GRF perturbation:
\begin{equation}
\psi(x,y) = \psi_{\mathrm{smooth}}(x,y | b,q,\theta_{\mathrm{pa}},x_0,y_0,\gamma,\phi)+\delta\psi(x,y | \sigma_{\delta\psi},\beta),
\end{equation}
with seven free parameters from the smooth lens potential and two from the GRF. A new realization of the stochastic GRF, $\delta\psi$, is generated for each mock lens. An example of a smooth and perturbed lens potential is shown in ~\citet[][fig. 2]{Vernardos_2020}.

\begin{figure*}
\centering
\includegraphics[width=0.97\textwidth]{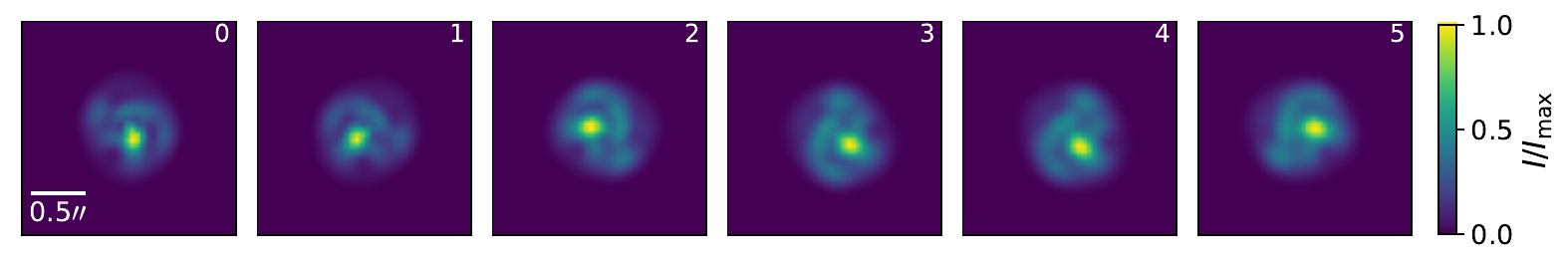} \\
\caption{Example data augmentation of the best-fitting source for lens J1621+3931 used to create data set 1. The leftmost panel is the nominal reconstructed source from~\citet{Shajib_2021}. The subsequent panels are continued augmentations of the nominal profile. To generate a training set, the nominal source is augmented at least once and up to five times. Each augmentation further degrades the image from the nominal source while maintaining its general shape and size. All panels have the same dimensions, i.e. $2''$ on a side with a resolution of $0.05''$ and $0.5''$ shown in the leftmost panel.}
\label{source_galaxy_example}
\end{figure*}

\begin{figure*}
\centering
\includegraphics[width=0.97\textwidth]{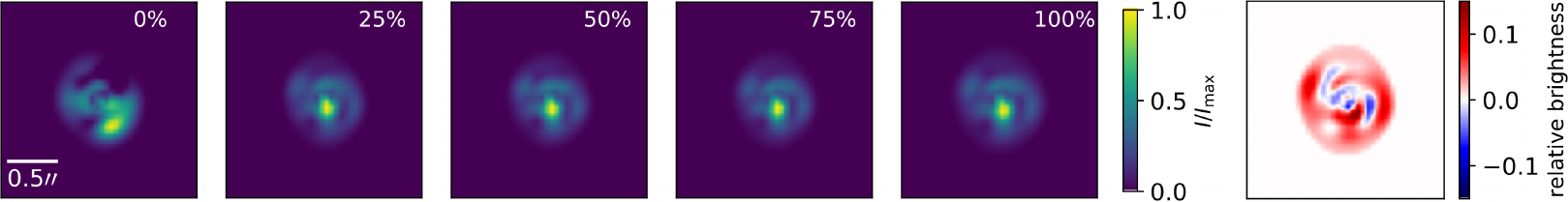}
\caption{The left five panels show different sources from the MCMC chain produced by~\citet{Shajib_2021} for lens J1621+3931 and used to create data set 2. The percentage through the MCMC chain is given in the top right of each panel where 0 per cent is the first source produced in the chain and 100 per cent is the last. The first source (leftmost panel) has not converged yet and significantly deviates from the later reconstructions, but the reconstruction quickly converges, as seen in the other panels. The rightmost panel shows the residuals between the 100 and 75 per cent sources from the MCMC chain. The two sources visually appearing similar, but there are residuals of order 10 per cent between them. All panels have the same dimensions, i.e. $2''$ on a side with a resolution of $0.05''$ and $0.5''$ shown in the leftmost panel.}
\label{source_galaxy_example_MCMC}
\end{figure*}

\subsection{Source brightness profile} \label{Source_galaxy_profiles_section}

We use the reconstructed sources of our sample of 23 SLACS lenses from~\citet{Shajib_2021} as source brightness profiles for our mock lenses. Details about their lens modelling technique can be found in section 3 of their paper, and the nominal source brightness profiles are shown in their figs. 2, 3, and 4. We focus on how they obtained their best-fitting sources as well as the full posterior distribution of the source parameters. The source galaxy brightness profiles are modelled using a basis of shapelets~\citep{Shapelets} superimposed on an elliptical Sérsic profile (the elliptical Sérsic profile is described in Section~\ref{lens_galaxy_light_sec}). Shapelets are a complete set of orthogonal basis functions, and a brightness profile can be fit with a finite subset of the basis functions by optimizing the shapelet coefficients. This parameterization reduces the number of free source parameters by at least an order of magnitude (from thousands to less than a hundred) compared to using a pixelated grid~\citep{Lens_modeling_with_basis,Shapelets_for_lense_modeling}. Shapelet based source reconstructions have previously been used to model quadruply lensed quasars~\citep{Used_Shapelets_Quasar} and time-delay lenses~\citep{Used_Shapelets_Hubble1,Used_Shapelets_Hubble2}. \citet{Shajib_2021} used MCMC sampling to obtain the posterior distributions of their model parameters, i.e. the lens mass parameters and the Sérsic and shapelet coefficients of the source.

We create two training sets with different methods of producing sources and a third training set that combines mocks from both methods. The first training set (data set 1) uses sources produced by common data augmentation operations applied to the 23 best-fitting sources from~\citet{Shajib_2021}. The data augmentation scheme used to generate sources for data set 1 includes the following steps: 
\begin{enumerate}[1.]
  \item Select one of the 23 reconstructed source brightness profiles, which have been normalized by their maximum brightness.
  \item Randomly select the number of times to augment the source between 1 and 5 times, all with equal probability.
  \item Multiply the source profile by a grid of correlated noise centred at one, with a standard deviation selected from the interval $[0, 0.1]$, and a correlation chosen from the range $[0.05'', 0.2'']$.
  \item Randomly flip, rotate the image by up to 360\degree, shear by up to 1\degree~clockwise or counterclockwise, and scale the image up or down by up to 2.5 per cent.
  \item Repeat steps 3 and 4 by the number of times selected in step 2.
\end{enumerate}
The correlated noise used for data augmentation is generating by convolving Gaussian white noise with a Gaussian kernel with standard deviation representing the length scale of the correlation, described in more detail in Section~\ref{noise_sec}. An example of the data augmentation procedure is given in Fig.~\ref{source_galaxy_example} for lens J1621+3931. After one augmentation the image remains very similar to the original best-fitting source (besides the rotation), however, with each subsequent augmentation it diverges further away. 

The second training set (data set 2) is created by including reconstructed sources from further down the MCMC chain. This way, while the algorithm is searching the parameter space for the best-fitting solution, but has not fully converged yet, the corresponding sources will appear ``perturbed'' but this time through a physical model (i.e. through the lens parameters of the mass model and the Sérsic and shaplet coefficients of the source model). We randomly select sources uniformly across the entire MCMC chain for each mock lens. We further diversify the training set by randomly flipping and rotating the images up to 360\degree~for each source in the same way as the first data set but without changing the shape of the source. Figure~\ref{source_galaxy_example_MCMC} shows examples of different sources obtained from the MCMC chain.

Our first method has the benefit of teaching the CNN to be invariant under small changes in the source morphology. In particular, the repeated multiplication of the nominal sources by correlated noise can change the power spectrum of the source images. We want to ensure the CNN is invariant to changes in the power spectrum of the sources to avoid confusing them with the power spectrum of the lensing perturbation. The downside is that we also include sources in our training set that can be far from the best-fitting. Our second method benefits from taking sources directly from the MCMC chain, which should be perturbed in a physically justifiable way and thus be more representative of the SLACS data. One potential caveat is that using only sources from the MCMC chain may systematically bias the training set, since the MCMC will not converge exactly to the true source brightness profile of the galaxy. We also create a third training set (the combined data set) by randomly selecting half of the prior two training sets. We later show in Section~\ref{Application_to_real_data_section} that the resulting predictions on the SLACS lenses are consistent across CNNs trained with each data set, so the two methods perform similarly.

\subsection{Lens galaxy light}
\label{lens_galaxy_light_sec}

In~\citet{Bayer_2023_1,Bayer_2023_2} the lens light was subtracted before estimating substructure properties, while~\citet{Vernardos_2020} used different masks assuming perfect lens light subtraction. We choose to explicitly include the lens light to avoid complications due to imperfect lens light subtraction.
We model the lens light as an elliptical Sérsic profile~\citep{Original_Sersic_Paper} given by:
\begin{equation}
I(x,y) = I_e \exp\left[-b_n\left\{\left(\frac{\sqrt{q^2 x\prime^2+y\prime^2}}{r_{\mathrm{eff}}}\right)^{1/n} - 1\right\}\right] ,
\end{equation}
where the parameters $\{q,\theta_{\mathrm{pa}},x_0,y_0\}$ are defined in the same way as the SIE potential, $r_{\mathrm{eff}}$ is the effective (or half-light) radius, $I_{\mathrm{e}}$ is the amplitude, and $n$ is the Sérsic index. The parameter $b_n$ is fixed such that the luminosity contained in $r_{\mathrm{eff}}$ is half the total. In real lensed systems, the lens light does not necessarily exactly coincide in location, shape, and orientation with the mass. We integrate this effect into our simulation by slightly deviating the lens light parameters from the lens mass parameters. This is achieved by drawing the lens light parameters randomly from a normal distribution centred at the lens mass parameters with a standard deviation equal to 2.5 per cent of the maximum range (0.0125 for $q$, $2.25\degree$ for $\theta_{\mathrm{pa}}$, and $0.005''$ for $x_0, y_0$).

Although the lens light of elliptical galaxies is typically close to a de Vaucouleurs' profile~\citep{Vaucouleurs_paper} with $n = 4$, we include a wide range of the Sérsic index, $n \in [3.0,8.0]$, to ensure that our choice will not bias the training set. The effective radius is correlated to the Einstein radius of the SIE deflection potential. In~\citet{Sersic_size}, they found that the effective radius scales approximately linearly with the Einstein radius with a best-fitting of $b = 0.576 \cdot r_{\mathrm{eff}}$ but with large deviations from the best-fitting line. We set the range of the effective radius to $r_{\mathrm{eff}} \in [0.5,2.6]\cdot b$ in order to encapsulate the deviations from the linear best-fitting. The lens light is added after the source galaxy is lensed in order to set the value of $I_e$ in such a way to closely mimic the contrast between the lensed source and lens light in the SLACS data. A log-ratio of $\log_{10}(R) \in [-0.1,2.0]$ between the maximum of the lens light and the lensed source is selected. The maximum ratio of $R = 100$ is justified by Fig.~\ref{real_data}, since the lensed source can always be seen at levels greater than $10^{-2}$. The minimum value is slightly less than $R = 1$ since the lens light should typically be brighter than the lensed source.

\subsection{Point spread function and noise}
\label{noise_sec}

\begin{figure*}
\centering
\includegraphics[width=0.97\textwidth]{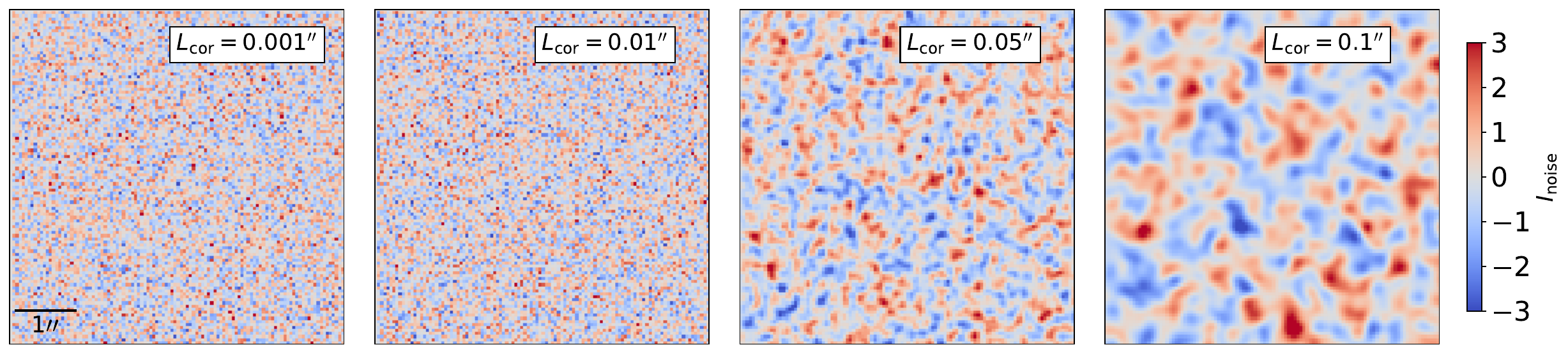}
\caption{Correlated noise for four different correlations $L_{\mathrm{cor}}$, each with a noise strength of $\sigma_{\mathrm{noise}} = 1$. All panels have the same dimensions, i.e. $5.5''$ on a side with a resolution of $0.05''$ and $1''$ shown in the leftmost panel.}
\label{correlated_noise}
\end{figure*}

After lensing the source and adding the lens light, we must convolve the image with a point spread function (PSF) and add noise to mimic the instrumental effects present in the observed data. Some of the SLACS data were taken in the F555W filter of the ACS instrument and some in F606W filter of the WFPC2. The PSFs for each of our 23 lenses were obtained by~\citet{Shajib_2021} using the \texttt{Tiny Tim} software\footnote{\href{https://github.com/spacetelescope/tinytim}{https://github.com/spacetelescope/tinytim}}~\citep{Tiny_Tim}. For each mock lens, we randomly select a PSF from the 23 available and convolve it with the lensed image. We randomly select the PSF instead of matching the PSF to each source to diversify the training set and force the network to distinguish between the instrumental effects and the lensed source.

The SLACS data were reduced through the drizzling procedure~\citep{Drizzling_correlated_noise}, which is known to induce correlations between neighbouring pixels~\citep{Drizzle_effect_Hubble}.
These correlations can additionally be enhanced by the charge-transfer inefficiency~\citep{CTI2,CTI1}.
Correlated noise was included in the dark matter subhalo study of~\citet{Dark_matter_CNN_2} who trained a CNN to detect dark matter subhaloes in lensed galaxies and found a 10--20 per cent loss in accuracy compared to using Gaussian white noise. \citet[][sect. 4.5]{Bayer_2023_1} describes a detailed procedure for obtaining the noise properties of drizzled \emph{HST} images using both the background sky noise and photon shot noise.

We introduce correlated noise into our simulated images using a simpler and more general approach. We first create a grid of uncorrelated Gaussian white noise at the same resolution as the data and then convolve it with a Gaussian kernel to induce a correlation. The standard deviation of the Gaussian kernel, denoted $L_{\mathrm{cor}}$, determines the noise correlation length. Figure~\ref{correlated_noise} shows examples of correlated noise for four different values of $L_{\mathrm{cor}}$. As $L_{\mathrm{cor}}$ increases, the noise gets more and more correlated between neighbouring pixels and appears clumpier. When $L_{\mathrm{cor}} = 0''$, we recover the white noise case where each pixel is completely uncorrelated.
We sample the correlation length uniformly in $L_{\mathrm{cor}}\in [0'', 0.05'']$ to induce Gaussian correlations with a standard deviation up to one pixel length. Incorporating correlated noise into our simulation is particularly important, as the noise power spectrum could interfere with measuring the power spectrum of the GRF perturbation. It is essential for our CNNs to learn to distinguish between these two effects to obtain unbiased results.

The Gaussian noise level $\sigma_{\mathrm{noise}}$ (i.e. the root mean squared deviation of the noise) for each of our lenses was estimated by~\citet{Shajib_2021}.
Most of the SLACS lenses have very low noise levels, with J073+3216 being the noisiest with $\sigma_{\mathrm{noise}} = 0.0048$ when the peak brightness is normalized to one. Most images have $\sigma_{\mathrm{noise}} < 0.002$. We sample the logarithm of the noise level uniformly in $\log_{10}(\sigma_{\mathrm{noise}}) \in [-3.25,-2.25]$ to encompass the SLACS data. We then create a realization of correlated noise for a given $\sigma_{\mathrm{noise}}$ and $L_{\mathrm{cor}}$, as described above, and add it to the lensed image. The images are normalized again by dividing by the peak brightness. This method of generating correlated noise is also used in our data augmentation scheme to generate source brightness profiles for data set 1, as described in Section~\ref{Source_galaxy_profiles_section}.

\subsection{Mock data set}
\label{create_data_set_sec}

The parameters required to generate a mock lens are listed in Table ~\ref{Parameter_table}.
Once a parameter vector is selected by uniformly drawing from the listed range, we use the Mock Lenses in Time software package\footnote{\href{https://github.com/gvernard/molet}{https://github.com/gvernard/molet}}~\citep[\texttt{MOLET};][]{MOLET} to simulate the images. Each data set consists of 250,000 lensed images with 230,000 used as the training set and 20,000 used as a validation set. We also include a final test set for each method with 25,000 images. The third combined data set uses a random half of data set 1 and 2. A random sample of 30 mock lensed images from the combined data set is shown in Appendix~\ref{Appendix_lens_images}.

\section{Convolutional neural network}
\label{CNN_section}
Our goal is to build and train a CNN to predict the GRF power spectrum parameters $\log_{10}(\sigma_{\delta\psi}^2)$ and $\beta$.
Uncertainty estimation plays a major role in our approach.
In standard regression problems, a neural network is trained using supervised learning to produce point estimates.
Only an overall systematic uncertainty can be estimated based on the network's performance on the test set using a metric such as the root mean squared error. 

We want to obtain uncertainties on each prediction instead of point estimates.
The uncertainty in ML can be broken up into the aleatoric and epistemic uncertainties.
The aleatoric uncertainty is the systematic and irreducible uncertainty that comes from the training data itself including factors such as the noise in the image, the impact of the GRF, the source galaxy morphology, among others. The epistemic uncertainty is the statistical uncertainty on the network parameters and was not included in the original work of~\citet{Vernardos_2020}.
We first describe our treatment of these two sources of uncertainty and then the CNN architecture, training, and performance.

\subsection{Uncertainty}
\label{uncertainty}

\subsubsection{Training labels and aleatoric uncertainty} 
\label{training_labels}

In typical image regression tasks, a CNN is trained using supervised learning where each image has a training label corresponding to the true value of the target parameters.
Instead, we want to produce a probability distribution of the target parameters that incorporates the uncertainty in the predictions.
This is done through regression-via-classification using the method introduced in~\citet{Vernardos_2020}, where the output of our CNN consists of $N_p = 100$ classes corresponding to the bins of a discrete probability distribution.
The training labels are uniform probability distributions that include the ground truth values, i.e. the true $\log_{10}(\sigma_{\delta\psi}^2)$ and $\beta$, though the ground truth is never shown directly to the CNN. 

\begin{figure}
\centering
\includegraphics[width=0.45\textwidth]{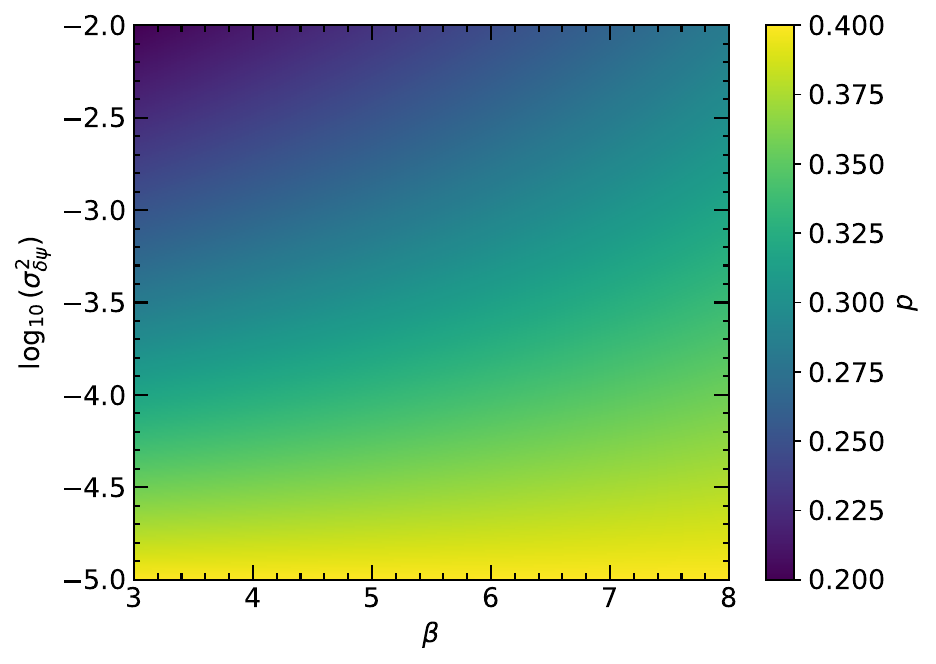}
\caption{Value of $p$ as a function of $\log_{10}(\sigma_{\delta\psi}^2)$ and $\beta$ from equation~(\ref{binomial_prob}). Notice the similarity with~\citet[][fig. 3]{Vernardos_2020}, but here without any dependence on the source or mass model.}
\label{p_with_A_and_beta}
\end{figure}

\begin{figure}
\centering
\includegraphics[width=0.47\textwidth]{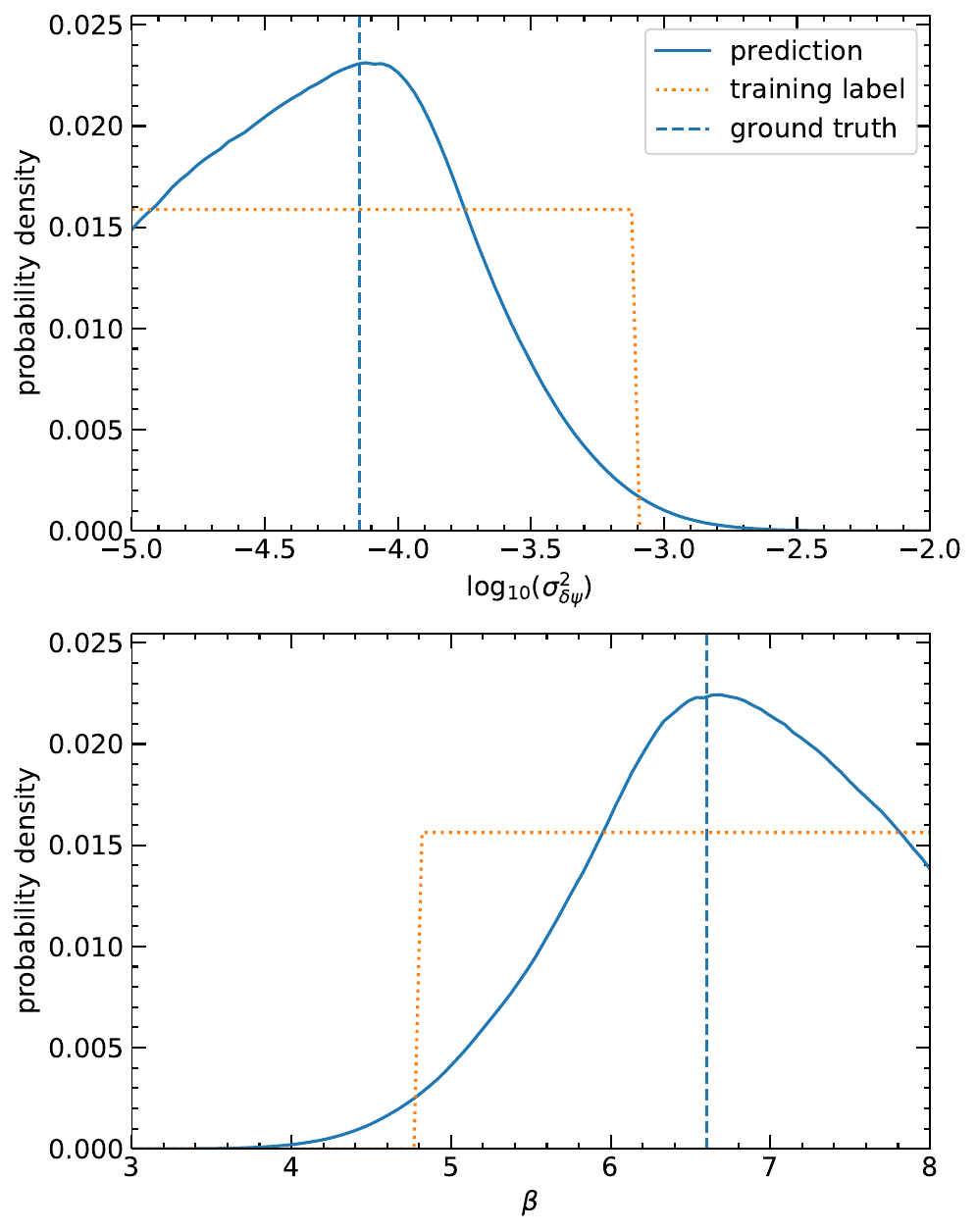}
\caption{Target distributions (dashed orange lines) and predictions of the CNN (blue solid lines) for $\log_{10}(\sigma_{\delta\psi}^2)$ (top) and $\beta$ (bottom) for one example mock image of the test set of the combined data set. The true value is indicated by the vertical dashed line but is never given directly to the CNN. Our predictions are obtained by drawing 200 samples from our BNN and averaging them. The first five samples are shown in Fig.~\ref{MC_figure}.}
\label{example_prediction}
\end{figure}

To create these target probability distributions,~\citet{Vernardos_2020} used the total power $P$ of residuals between perturbed and unperturbed lensed images. The larger the residuals, the more pronounced the perturbations and their imprint on the lens, thus receiving narrower target distributions. Conversely, smaller perturbations are more challenging to measure, and therefore, higher uncertainties should be associated with the input labels. In this work we choose an analytic expression that only depends on our GRF parameters $\log_{10}(\sigma_{\delta\psi}^2)$ and $\beta$.
This approach is more appropriate to apply to a larger lens sample since it is independent of our choice of source and mass model, which can vary greatly from lens to lens. Consistent with~\citet{Vernardos_2020}, we start from the ground truth and assign a range around either side of it, determined by the number of classes away form the ground truth. Each range is independently drawn from a binomial distribution with parameters $N$ and $p$ given by:
\begin{eqnarray}
    N & = & n \, N_{\mathrm{p}} , \nonumber \\
    p & = & p_0 r^{-\frac{\log_{10}(\sigma_{\delta\psi}^2)-\log_{10}(\sigma_{\mathrm{min}}^2)}{\log_{10}(\sigma_{\mathrm{max}}^2)-\log_{10}(\sigma_{\mathrm{min}}^2)}\cdot \left(1-\frac{1}{2}\frac{\beta-\beta_{\mathrm{min}}}{\beta_{\mathrm{max}}-\beta_{\mathrm{min}}}\right)} ,
 \label{binomial_prob}
\end{eqnarray}
where we set $p_0 = 0.4$, $r = 2.0$, and $n = 0.9$. Fig.~\ref{p_with_A_and_beta} shows how $p$ varies across the parameter space of $\log_{10}(\sigma_{\delta\psi}^2)$ and $\beta$, matching the behaviour of \citet[][fig. 3]{Vernardos_2020}. Equation~(\ref{binomial_prob}) is chosen to behave similarly to~\citet[][equation 4, 5]{Vernardos_2020} but only depend on the GRF power-law parameters.  
In~\citet{Vernardos_2020} $n=0.6$ was arbitrarily used, while here we use a larger value to avoid underestimating the uncertainty (further discussed in Section~\ref{Training}). By design, the ground truth value could be anywhere in the uniform probability distributions of the training labels but tends to closer to the centre. Figure~\ref{example_prediction} shows an example target and predicted distribution of our fully trained CNN for one mock image in our test set. After every epoch the training labels are regenerated, drawing new values from the binomial distribution each time.

\subsubsection{Epistemic uncertainty}
\label{Epistemic_uncertainty}

\begin{figure}
\centering
\includegraphics[width=0.47\textwidth]{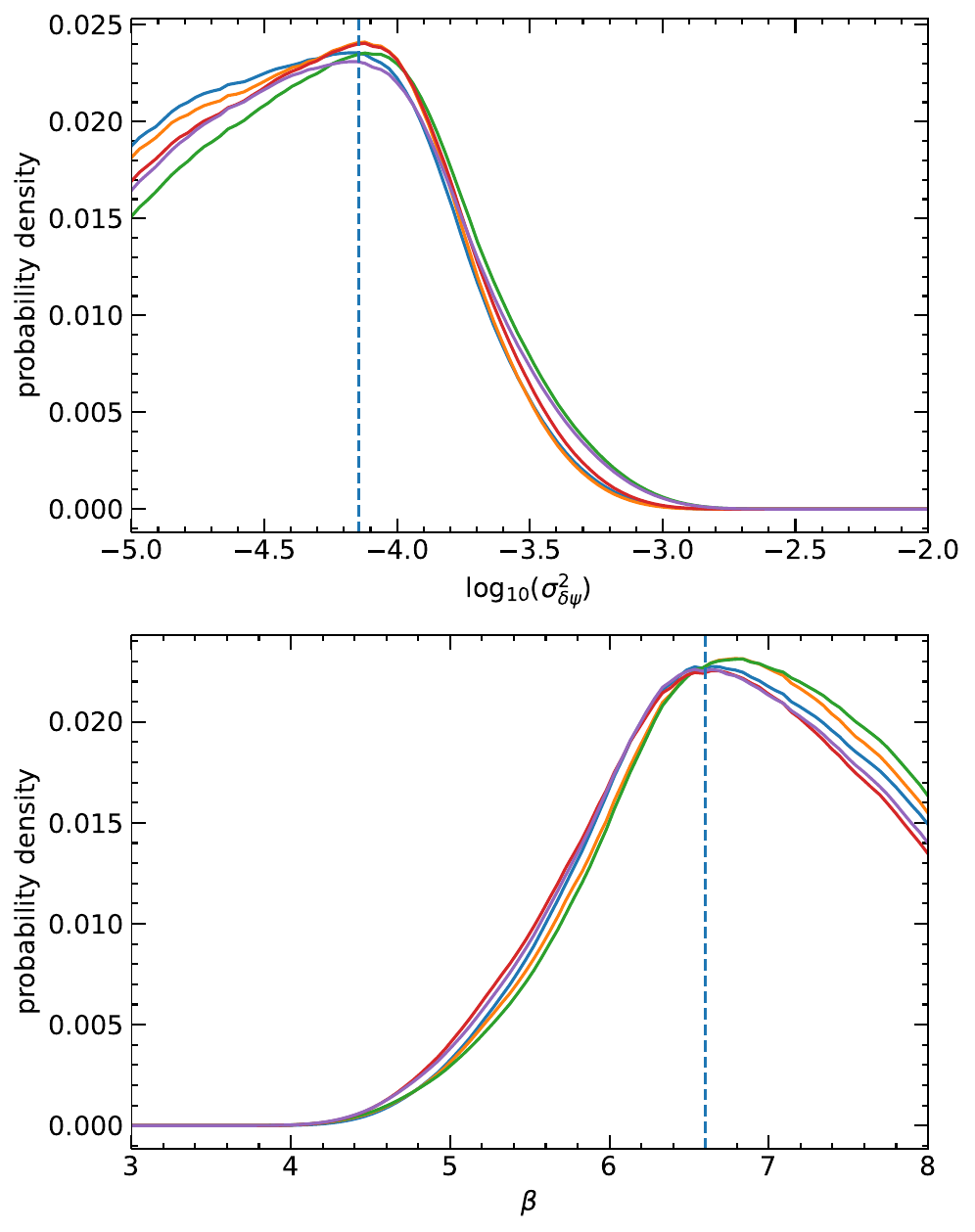}
\caption{Example of multiple predictions made by our BNN for the same test set example as in Fig.~\ref{example_prediction}. The first five samples are shown in different colours for $\log_{10}(\sigma_{\delta\psi}^2)$ (top) and $\beta$ (bottom). The true value is indicated by the dashed vertical line. We include the epistemic uncertainty by averaging across the different samples, increase the overall uncertainty by widening the probability distributions.}
\label{MC_figure}
\end{figure}

We use a Bayesian neural network (BNN) to obtain uncertainty in the model parameters. Each time the network makes a prediction, a random value of each weight is drawn from a Gaussian probability distribution with trainable mean and variance rather than having fixed weights across the network~\citep{VI}. During inference, we repeatedly sample from the network to obtain an ensemble of different predictions. The epistemic uncertainty is included by averaging across the ensemble of 200 predictions from the BNN. Figure~\ref{MC_figure} shows five predictions of the network for the same test set example as in Fig.~\ref{example_prediction}. The epistemic uncertainty can be represented as the increase in the uncertainty due to averaging across the different samples. 

Several methods are available for efficiently sampling from the weight distributions of a BNN. We use the most common one, Flipout~\citep{Flipout}, an efficient technique for optimizing a BNN by generating pseudo-independent weight perturbations on mini-batches of data. Unlike traditional methods where weight samples are shared across all examples in a mini-batch, Flipout applies a random sign matrix to the weights for each data point. This approach efficiently decorrelates the gradients between different examples in the mini-batch, reducing the gradient variance compared with shared perturbations and thereby stabilizing the training process. \citet{CMB_Flipout1} showed that Flipout outperforms other methods of sampling from the weight distributions of a BNN during training, including Dropout, DropConnect, and the reparameterization trick. The log-variance of the kernel posterior is initialized by $\mathcal{N}(-9,0.01)$ consistent with~\citet{CMB_Flipout1}. In variational inference, the negative evidence lower bound object (ELBO) is minimized with respect to the variational parameters~\citep{VI,MC_Dropout,Uncertainty_in_SIE_predictions}. The Kullback–Leibler (KL) divergence between the variational distribution and the prior is included in our loss to act as a regularization term to control the standard deviation of each weight in the network. We also use KL-annealing~\citep{KL_annealing} linearly for the first 100 epochs to slowly introduce the KL-divergence term. The authors of~\citet{CMB_Flipout1} discuss how $L_2$ regularization can be used on the standard deviation of each weight to calibrate the uncertainty in our predictions. We apply an $L_2$ regularization of $10^{-3}$ on the standard deviation. Lowering the regularization leads to higher uncertainties on each weight, while increasing the regularization makes the network behave more deterministic by pushing the variances closer to zero. This parameter can be adjusted in the same way that the dropout rate can be adjusted using Monte Carlo dropout~\citep{MC_Dropout} to calibrate the uncertainty~\citep{Uncertainty_in_SIE_predictions}. We note that the BNN will have twice as many parameters as a corresponding deterministic network because we optimize both the mean and standard deviation of each weight.

\subsection{Convolutional neural network architecture}
\label{architecture}
The CNN architecture used in~\citet{Vernardos_2020} consisted of a series of 7 convolutional layers that were then flattened and followed by 2 fully connected layers for each parameter.
In this work we use a variant of ResNet-18~\citep{Resnet2015} to include residual skip connections into our CNN. The addition of skip connections enables CNNs to be significantly deeper without suffering from vanishing gradient and degradation issues and has been shown to improve gradient flow and training stability. We also use squeeze and excitation connections~\citep{Squeeze_and_excitation} to add a spatial attention mechanism to our network. The input of the CNN is $110 \times 110$ pixel images ($5.5''$ across with a resolution of $0.05''$ per pixel to match the SLACS data). The standard ResNet-18 starts with one convolutional layer with a filter size of 64 and 4 residual blocks with 4 convolutional layers each and filter sizes of [64,~128,~256,~512] respectively.
We use the same architecture except with smaller filter sizes of [32,~64,~128,~256] to use fewer parameters and avoid overfitting, and we include the addition of the squeeze and excitation connections with a ratio of eight~\citep{Resnet2015,Squeeze_and_excitation}.
The output of the final convolutional layer is flattened and split into two separate fully connected layers with $N_p = 100$ classes, representing the bins of the discrete probability distributions of our output in $\log_{10}(\sigma_{\delta\psi}^2)$ and $\beta$.
The final fully connected layers are deterministic instead of Bayesian since this has been shown to improve performance~\citep{CMB_Flipout1}.
We also includes a softmax activation function to normalize the final outputs to a probability distributions such that all classes are positive and sum to one. Our model\footnote{\url{https://github.com/JFagin/GRF_ML}} has 5,735,456 trainable parameters that are initialized randomly and is implemented using \texttt{TensorFlow} and the \texttt{TensorFlow-Probability} API~\citep{tensorflow2015}.

\subsection{Training and performance}
\label{Training}
\begin{figure}
\centering
\includegraphics[width=0.47\textwidth]{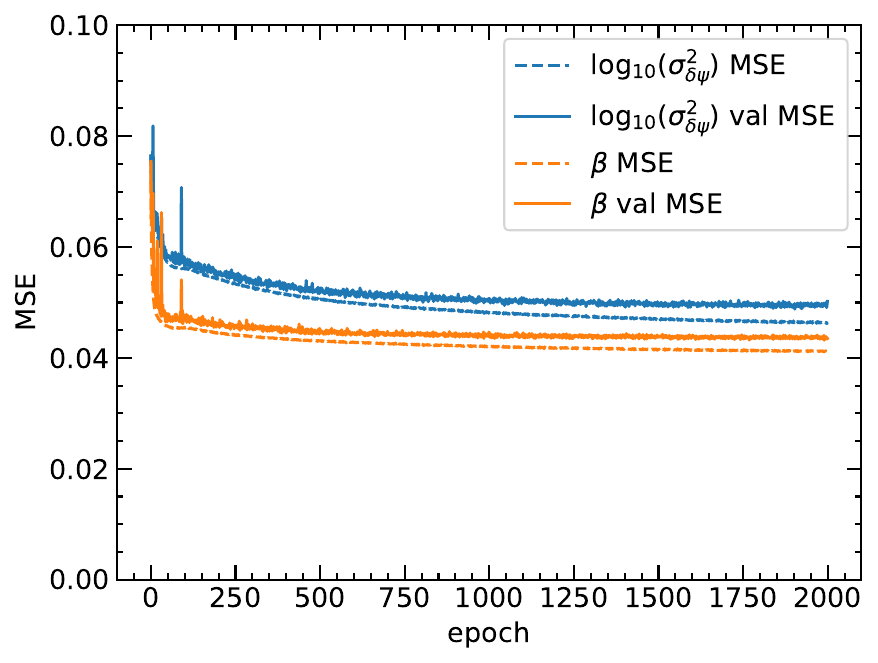}
\caption{Mean squared error as a function of epoch between the mean of each predicted probability distribution and the ground truth across the validation set of the combined data set. The ranges of our parameters are scaled from zero to one for comparison.}
\label{loss_vs_epoch}
\end{figure}

For our loss function we follow the approach of~\citet{Vernardos_2020} and minimize a Jensen-Shannon (JS) divergence term between the target probability distributions $Q$ and predicted probability distributions $P$ and an entropy term. The loss function is a weighted sum of the two terms and given by:
\begin{equation}
\mathcal{L}(P,Q) = \lambda \mathrm{JS}(P,Q)+(1-\lambda)H(P) ,
\label{Loss_function}
\end{equation}
where $\lambda$ determines the relative impact of each term. The JS-divergence is a smooth and symmetric version of the KL-divergence which minimizes the difference between the predicted probability distributions and the target distributions. The entropy term serves as a regularization term that only depends on the predicted probability distribution. Minimizing the entropy term alone is a non-parametric way of obtaining the smallest possible support for the predicted distribution.

In~\citet{Vernardos_2020} they used a value of $\lambda = 0.9$, but we use a value of $\lambda = 0.975$, which leads to the entropy term contributing $\approx 45$ per cent of the relative contribution to the loss. Our value for $\lambda$ is purposely closer to 1 in order to avoid learning narrower predicted probability distributions and thus underestimating the uncertainty. In addition to equation~(\ref{Loss_function}), the regularization term of the BNN described in Section~\ref{Epistemic_uncertainty} is appended to the loss to control the standard deviation of the model weights.

We train the CNNs through multiple passes of the entire training sets, referred to as epochs. Each CNN is trained for a total of 2,000 epochs. During both training and inference, we augment each image by randomly rotating in intervals of $90\degree$ and flipping vertically or horizontally. This teaches the CNNs to remain invariant under these transformations and effectively increases the number of training examples by a factor of 8, preventing overfitting. We find that augmenting the data during inference improves uncertainty estimates as we average the predictions across various orientations. We also sample a new realization of the training labels each epoch, as mentioned in Section~\ref{training_labels}.

Our CNNs are trained by minimizing the loss function using an Adam optimizer~\citep{Adam} with an initial learning rate of $10^{-3}$ and a batch size of 250. The learning rate is exponentially decayed to $10^{-4}$ over the course of training. Figure~\ref{loss_vs_epoch} shows how the mean squared error (MSE) between the mean of each predicted probability distribution and the ground truth varies with epoch across the validation set of the combined data set. We note that the MSE is indirectly minimized because the CNNs are never shown the ground truth values directly. The parameter space is scaled from zero to one for comparison, making the root MSE effectively represent a percentage error relative to the total parameter space. The validation MSE levels off by 2,000 epochs, so further training could lead to overfitting and be detrimental to the model performance. After the CNNs are trained, we apply them to the test sets to evaluate their performance. We report a final root MSE of 20.6 per cent for $\log_{10}(\sigma_{\delta\psi}^2)$ and 19.9 per cent for $\beta$ using the combined data set. We estimate the relative contribution of the epistemic and aleatoric uncertainties by comparing the standard deviation of our probability distributions for a single prediction and the average of 200 predictions for each test set example. We find that the standard deviation increased by 9.2 per cent for $\log_{10}(\sigma_{\delta\psi}^2)$ and 5.4 per cent for $\beta$ for the combined data set, which represents the percentage contribution of the epistemic uncertainty as compared to the aleatoric uncertainty. We expect the contribution of the epistemic uncertainty to be small compared to the aleatoric uncertainty since it should decrease with the size of the training set, and we use a relatively large training set of 220,000 images. Additional metrics and figures for data set 1 and 2 are given in Appendix~\ref{appendix_extra_metrics} and show similar performance.

\begin{figure}
\centering
\includegraphics[width=0.47\textwidth]{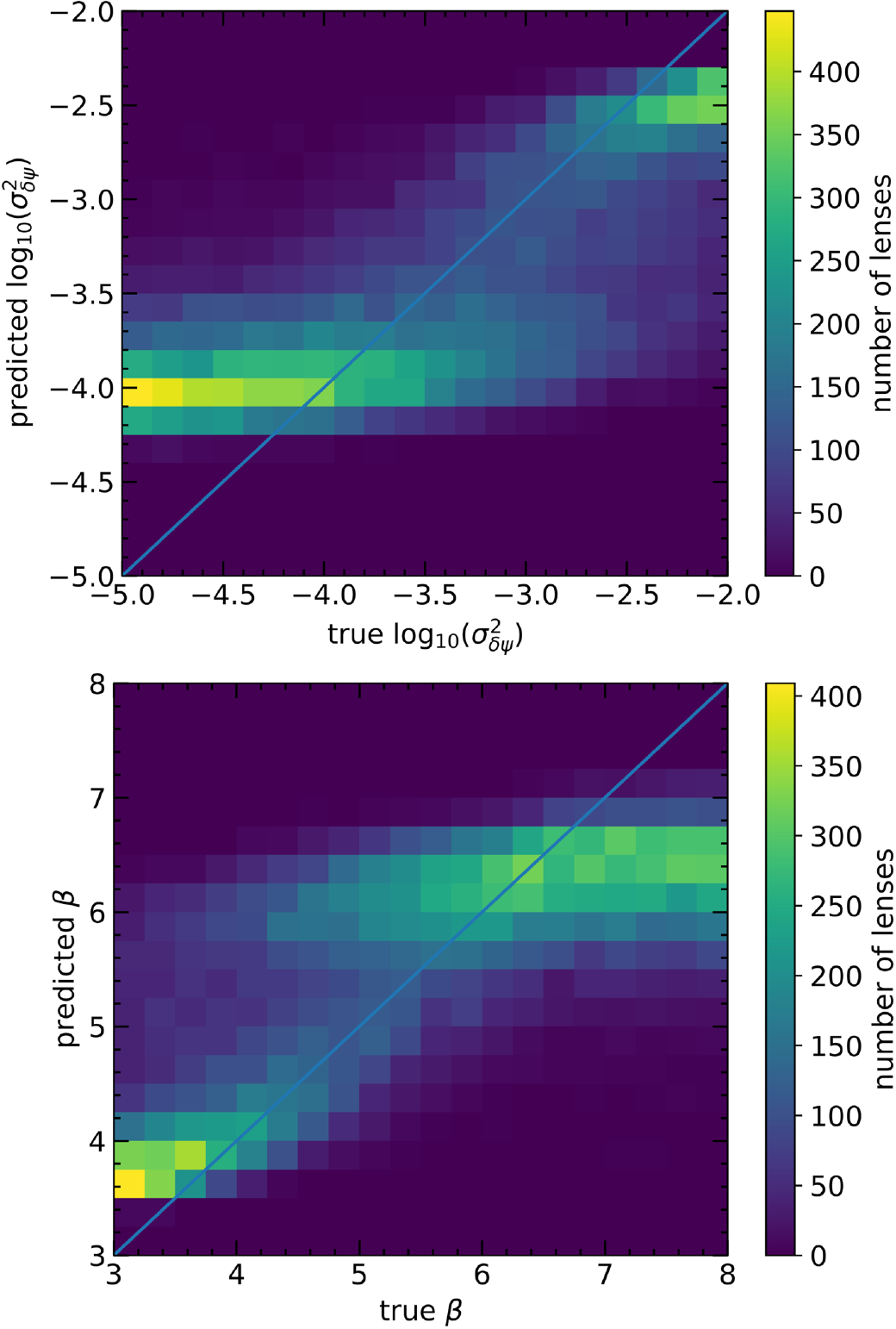}
\caption{Mean of our predicted probability distributions for $\log_{10}(\sigma_{\delta\psi}^2)$ (top) and $\beta$ (bottom) compared to the ground truth across the test set of the combined data set. The diagonal blue line represents the ideal case where each mean prediction would match the true value.}
\label{confusion_matrices}
\end{figure}

\begin{figure}
\centering
\includegraphics[width=0.47\textwidth]{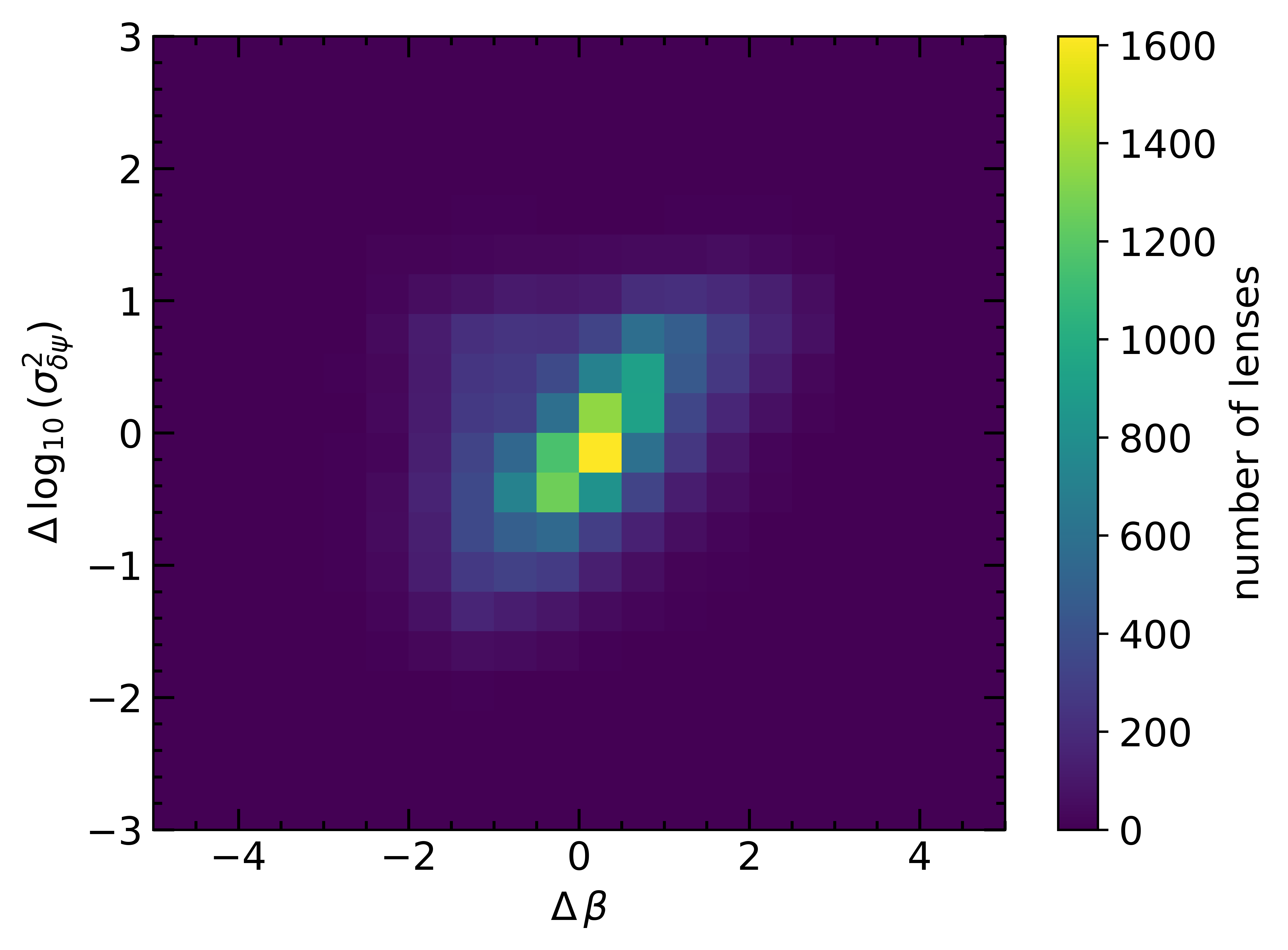}
\caption{Histogram of the difference between the mean of our predicted probability distributions and the ground truth across the test set of the combined data set. The ideal case where the mean prediction matches the ground truth occurs at the centre, i.e. when $\Delta \, \log_{10}(\sigma_{\delta\psi}^2)$ and $\Delta \, \beta$ are zero.}
\label{difference_hist}
\end{figure}

\begin{figure}
\centering
\includegraphics[width=0.47\textwidth]{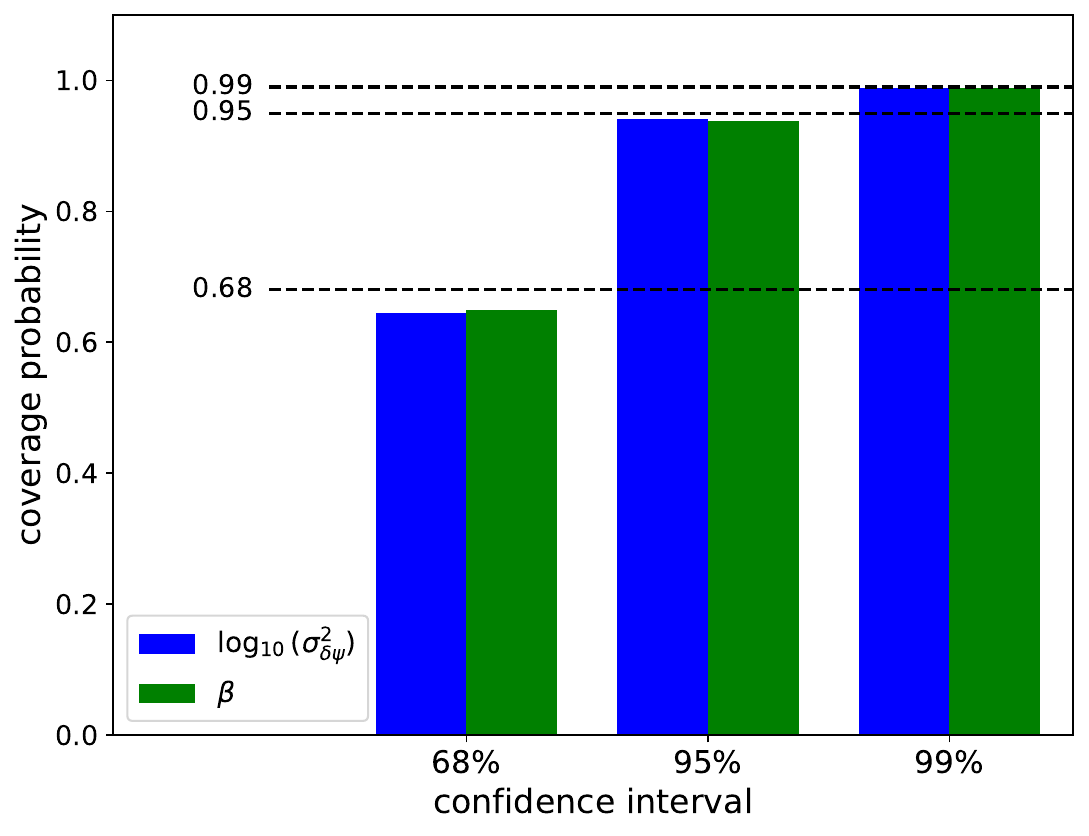}
\caption{Coverage probabilities for the combined data set, i.e. the fraction of the test set which has ground truth values within 68, 95, and 99 per cent confidence intervals. The ideal case of perfect uncertainty calibration, when the coverage probabilities exactly match the middle confidence intervals, is shown by the horizontal dashed lines.}
\label{convergent_probability}
\end{figure}

Figure~\ref{confusion_matrices} shows the mean of our predicted probability distributions compared to the ground truth. Ideally the mean predictions would equal the ground truth and follow the diagonal blue line. In many cases, our predictions include substantial uncertainties causing deviations between the mean predictions and the ground truth. By design, the mean of the output probability distributions can never reach the edges of the parameter space. That would be the case only if the distributions became delta functions, which is prevented by the JS-divergence term in the loss function (see equation~(\ref{Loss_function})).
Hence, the edges of the predictions are empty and we get a cluster of predictions on either side of the parameter space. While alternative metrics such as the mode could circumvent this boundary constraint, we find the performance to be inferior compared to using the mean or median. When using the prediction of our CNNs, the full probability distribution should be used instead of summary statistics to avoid bias. We also show a histogram of the difference between the mean of our predicted probability distributions and the ground truth for $\log_{10}(\sigma_{\delta\psi}^2)$ and $\beta$ in Fig.~\ref{difference_hist}. There is a correlation between the residuals in $\log_{10}(\sigma_{\delta\psi}^2)$ and $\beta$ since it is easier to predict $\beta$ when $\log_{10}(\sigma_{\delta\psi}^2)$ is large, as the lens perturbation is more impactful. 

The coverage probabilities are the fraction of ground truth values that lie within a given confidence interval of our predicted probability distributions across the test set. To assess the validity of our predicted probability distributions as uncertainties, Fig.~\ref{convergent_probability} shows the coverage probabilities for the 68, 95, and 99 per cent confidence intervals. We assess these confidence intervals since they represent the $1\sigma$, $2\sigma$, and  $3\sigma$ confidence intervals of a Gaussian, although our probability distributions can take on any form. Ideally, all the ground truth values would be within our reported confidence intervals at the same rate as our predictions. The converge probabilities nearly line up with the ideal case, so we can conclude that our predicted probability distributions approximately represent the uncertainty in our predictions. For better calibration of our uncertainties, we set $n=0.9$ in equation~(\ref{binomial_prob}) and $\lambda = 0.975$ in equation~(\ref{Loss_function}). 

\section{Application to SLACS lens sample}
\label{Application_to_real_data_section}

\begin{figure*}
\centering
\includegraphics[width=0.97\textwidth]{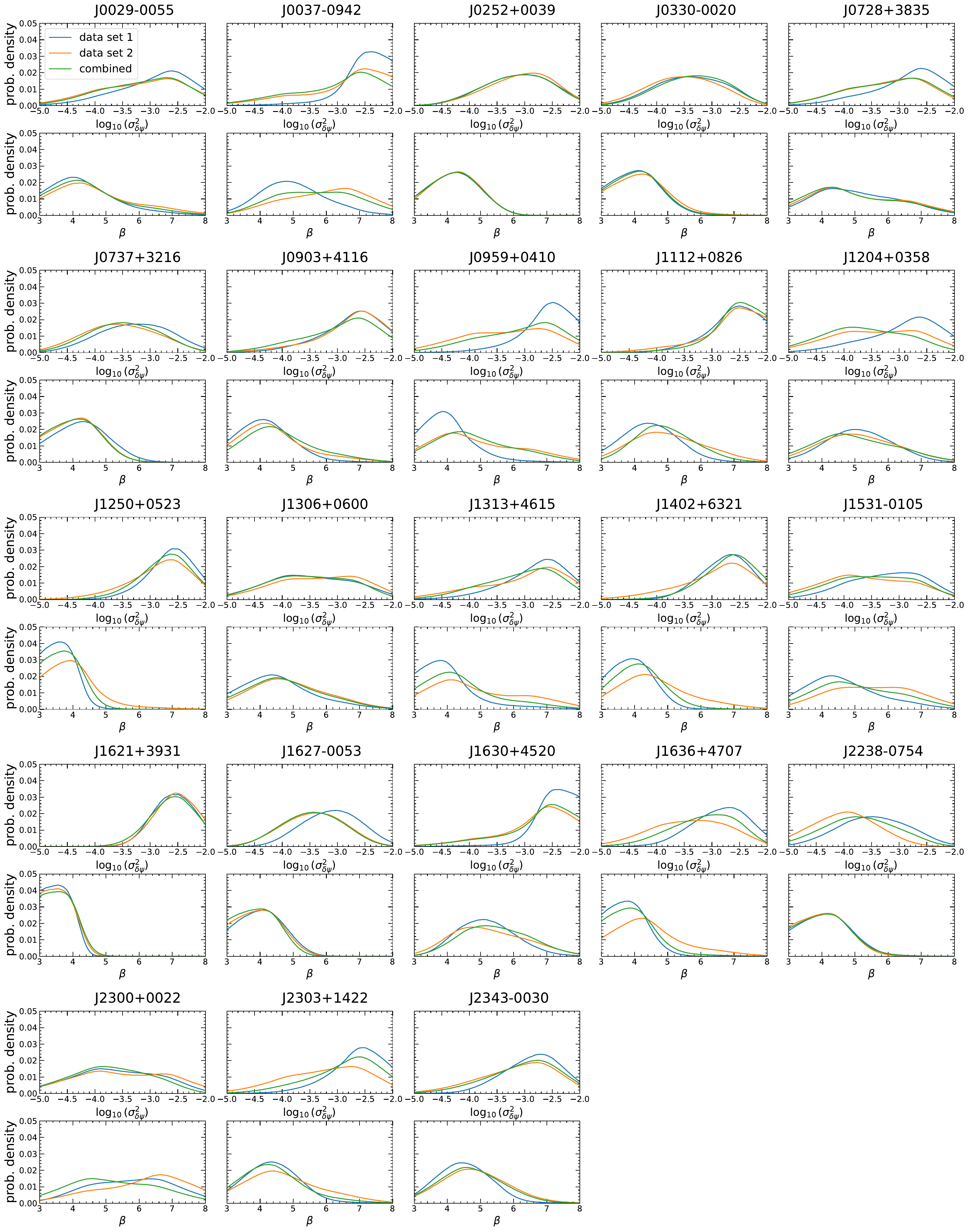}
\caption{Predicted probability distributions of $\log_{10}(
\sigma_{\delta\psi}^2)$ and $\beta$ for each of the 23 SLACS lenses from the CNNs trained with data set 1 (blue), data set 2 (orange), and the combined data set (green). Each probability distribution represents the combined aleatoric and epistemic uncertainties by averaging over 200 predictions of our Bayesian neural network. The average probability distributions across the three CNNs' predictions are given in Appendix~\ref{appendix_extra_metrics}.}
\label{probability_dist_of_data}
\end{figure*}

\begin{figure*}
\centering
\includegraphics[width=0.97\textwidth]{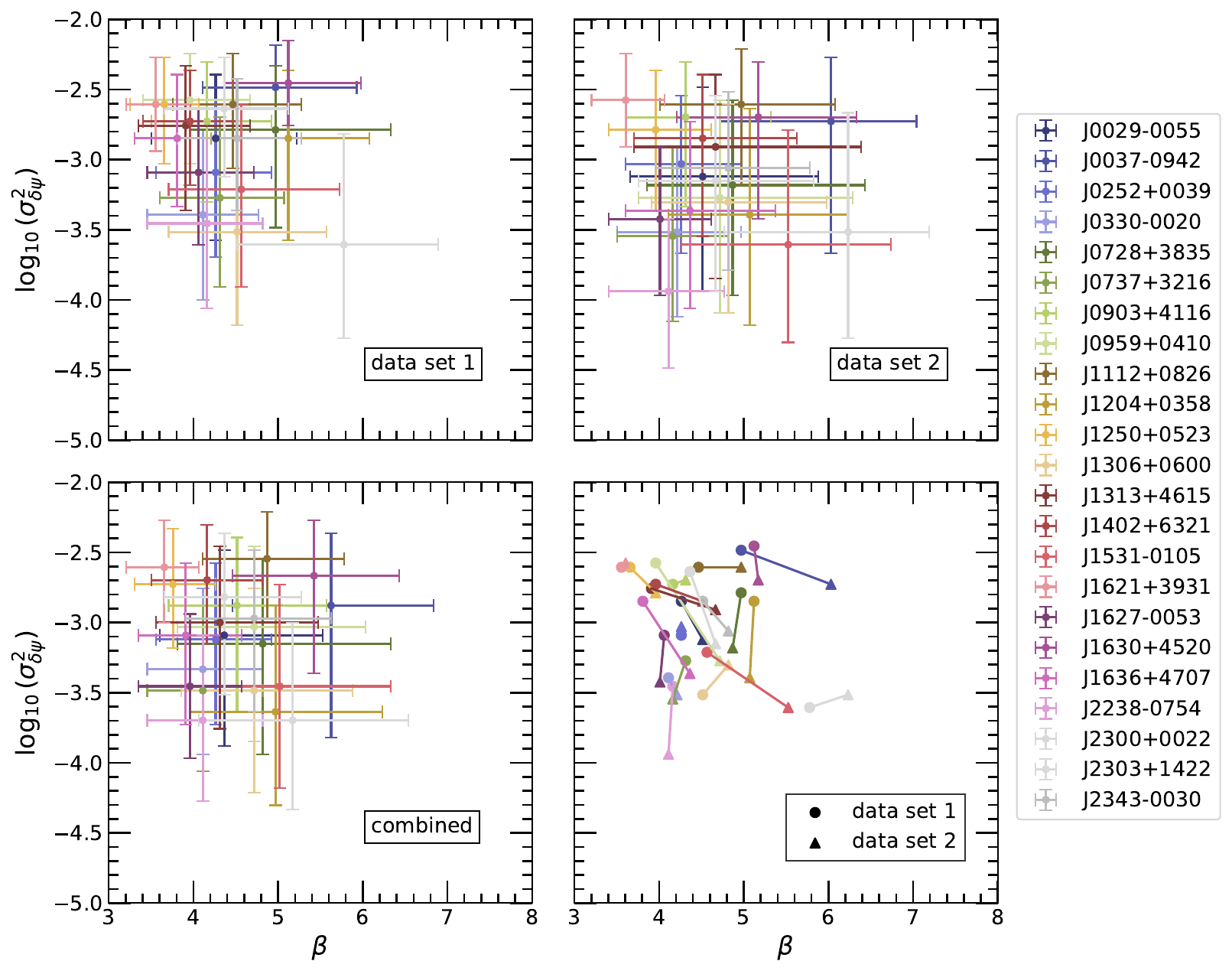}
\caption{The top two and bottom left panels show the median predictions and middle 68 per cent confidence intervals for our three separately trained CNNs. The bottom right panel shows how the median predictions change between using data set 1 (circles) and 2 (triangles). The numerical values of each prediction are given in Table~\ref{Table_measured_values}.}
\label{compare_differences_between_sources}
\end{figure*}

\begin{figure*}
\centering
\includegraphics[width=0.95\textwidth]{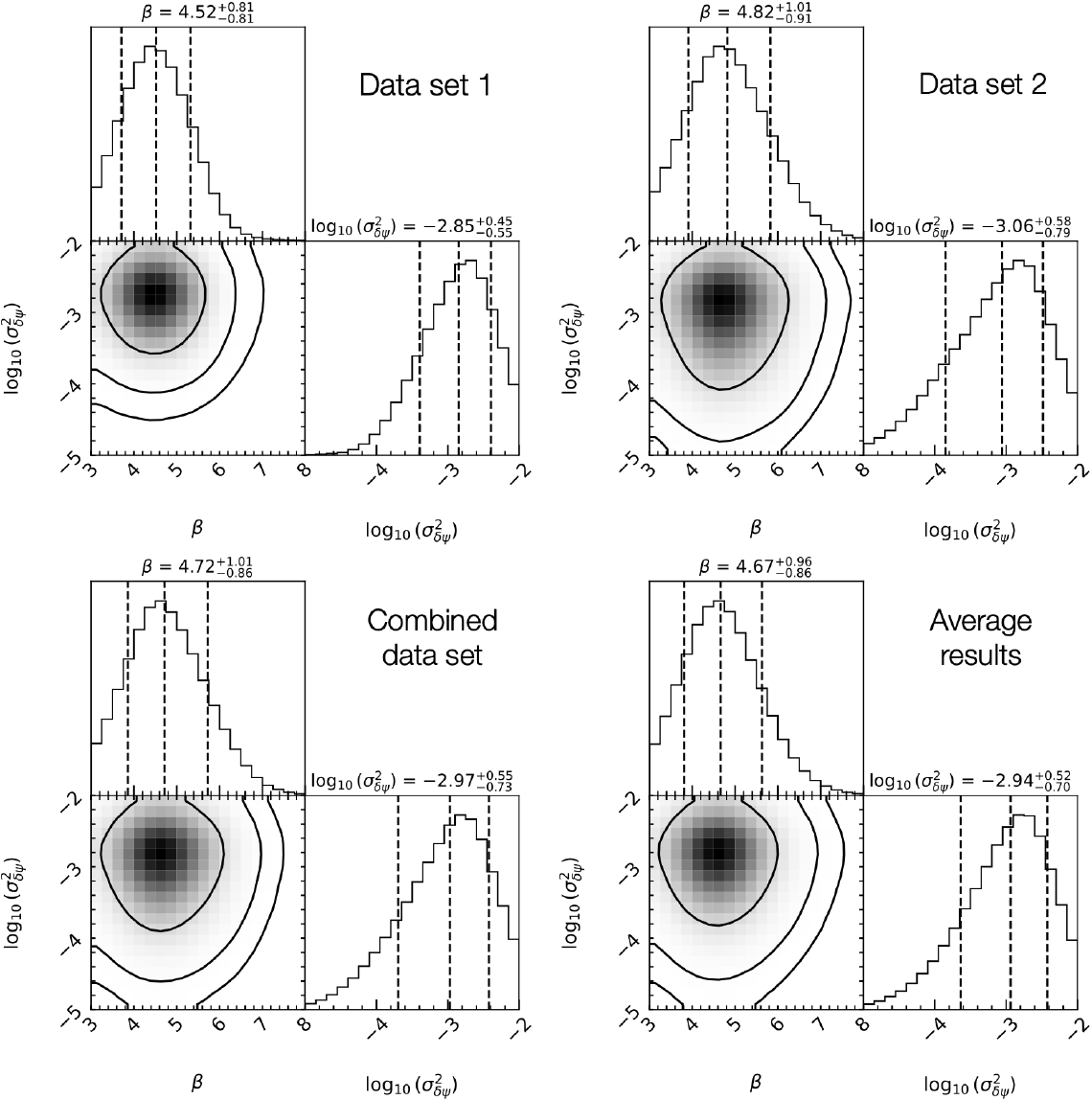}
\caption{Joint probability distributions across the 23 SLACS lenses for the prediction of the CNNs trained on each data set. The average between the three CNNs' predictions is given in the bottom right panel. The joint probabilities are obtained by multiplying the predictions of individual lenses across our SLACS lens sample. Each of the contours contain the 68, 95, and 99 per cent probabilities. The dashed lines in the marginal distributions indicate the median value and 68 per cent confidence interval, which are also given numerically on top.}
\label{real_data_average_prob_individual}
\end{figure*}

After our three CNNs are trained, we apply them to the SLACS lenses. The resulting probability distributions for $\log_{10}(\sigma_{\delta\psi}^2)$ and $\beta$ are given in Fig.~\ref{probability_dist_of_data}. While there are some differences between the three predictions across the lenses, they generally follow the same trends. This reassures us that the results have some resilience to how the training sets are generated between the different source methods. We include some additional systematic uncertainty, coming from applying the three different CNNs to the real data, by averaging across the separate results for each lens. The average predicted probability distributions are given in Appendix~\ref{appendix_extra_metrics}. Averaging across the results of separately trained models is a strategy known as ensemble learning~\citep[see][for a review of ensemble deep learning]{Ganaie_2022}. By combining the results of our separately trained CNNs, we can achieve better generalization on the real data and marginalize across the different methods of choosing sources for our lensing simulation. This averaging induces a conservative estimate of the uncertainty (i.e. wider confidence levels), since averaging across multiple predictions increases the uncertainty by widening our predicted probability distributions.

The median values and their uncertainties for each lens are shown in Fig.~\ref{compare_differences_between_sources}. We also show the shift in the median predictions of each individual lens between the CNN trained on data set 1 compared to data set 2 in the bottom right panel. This demonstrates that the median predictions remain relatively stable despite using different sources. The median values with a 68 per cent confidence interval are also given numerically in Table~\ref{Table_measured_values}. Our primary predictions can be taken as the average results across the predictions of our three CNNs.

Figure~\ref{real_data_average_prob_individual} shows the joint probability distributions across our SLACS lens population for each CNN. Since each observation is independent, we obtain the joint probability by multiplying the individual predictions together across the lens population. We then average the predictions of the separately trained CNNs to obtain a final joint probability distribution given in the bottom right panel. The overall trend is consistent between the three methods with average predicted power-law parameters $\log_{10}(\sigma_{\delta\psi}^2) = -2.94^{+0.52}_{-0.70}$ and $\beta = 4.67^{+0.96}_{-0.86}$ at the 68 per cent confidence level.

\begin{figure*}
\centering
\includegraphics[width=0.97\textwidth]{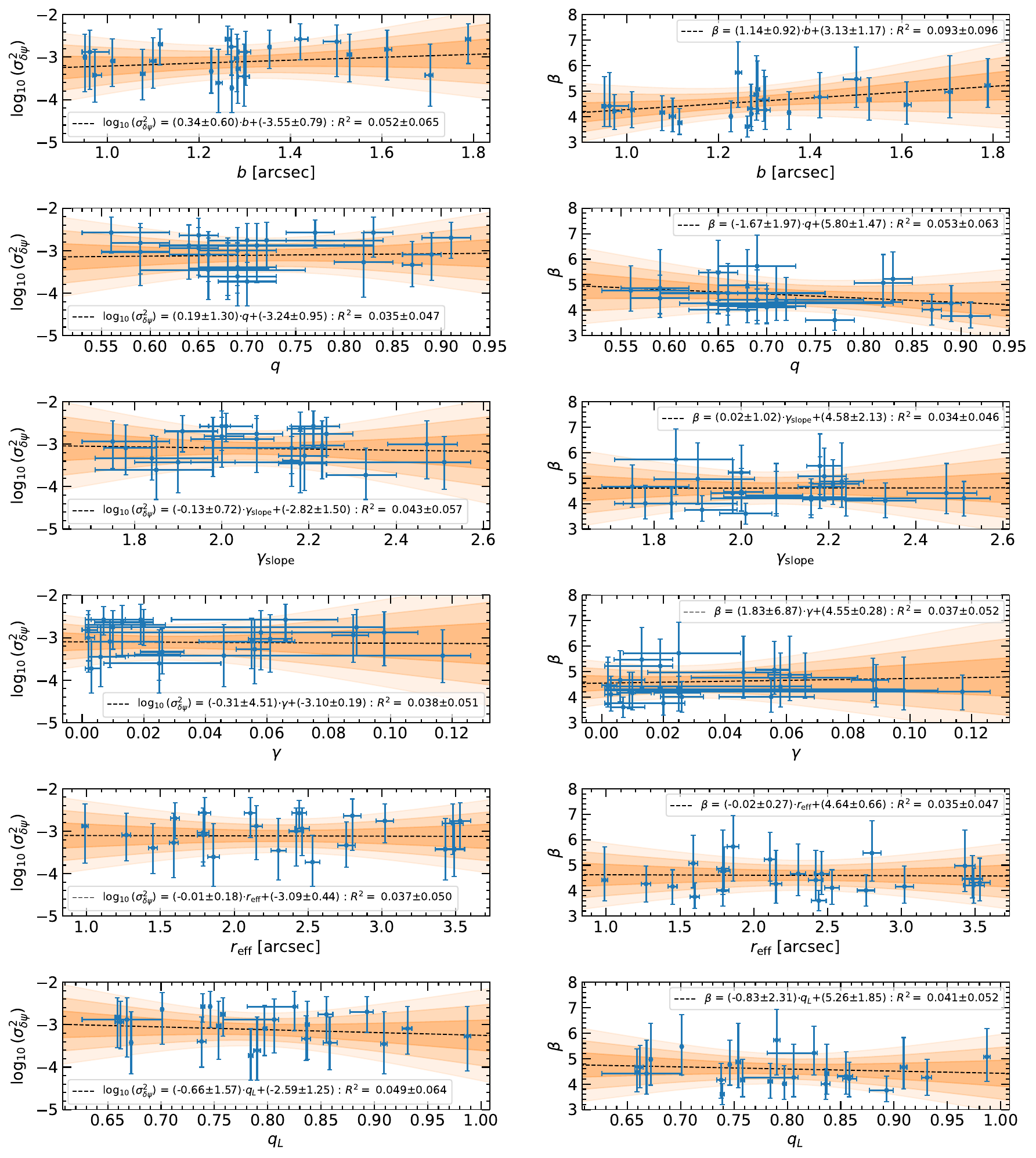}
\caption{Predicted $\log_{10}(\sigma_{\delta\psi}^2)$ and $\beta$ using the average predictions across our three CNNs compared to the lens mass and lens light parameters for the 23 SLACS lenses. The median predictions and $68$ per cent confidence interval are given in blue. The various lens mass and lens light parameters were estimated by~\citet[][table 1]{Shajib_2021}. Here $\gamma_{\mathrm{slope}}$ is the logarithmic slope of the mass profile and $q_L$ is the major to minor axis ratio of the lens light. We fit 10,000 linear best-fitting lines to the data by drawing values from our predicted probability distributions and lens model parameters to include the uncertainty in our fits. The mean linear fits are shown as the black dashed lines and the $1\sigma$, $2\sigma$, and $3\sigma$ uncertainties are given in orange. The best-fitting slope, intercept, and $R^2$ are given in the legend with uncertainties coming from the standard deviation across the ensemble of linear fits.}
\label{A_and_beta_vs_lens_parameters_avg}
\end{figure*}

Figure~\ref{A_and_beta_vs_lens_parameters_avg} shows our predicted power-law parameters averaged across our three CNNs compared to the smooth lensing parameters estimated by~\citet[][table 1]{Shajib_2021}. The same figures for the CNNs trained with data set 1, data set 2, and the combined data set are given in Appendix~\ref{appendix_extra_metrics} and show very similar trends. We find no strong correlation between our predicted substructure parameters and the smooth lens parameters. This indicates there is no substantial systematic bias in our results pertaining to the smooth lens parameters. The uncertainty on these trends is large with the relatively small lens population we analyse in this work, so more lenses may be required to definitively measure trends in the substructure and smooth mass parameters.

\begin{table*}
\centering
 \caption{Median predicted $\log_{10}(\sigma_{\delta\psi}^2)$ and $\beta$ and 68 per cent confidence intervals for our selected SLACS lenses using the CNNs trained with data set 1, data set 2, the combined data set, and then the average results across our three predictions. 
 }
 \begin{tabular}{ l c c c c c c c c} 
 & \multicolumn{2}{c}{data set 1} & \multicolumn{2}{c}{data set 2} & \multicolumn{2}{c}{combined data set} & \multicolumn{2}{c}{average result}   \\ 
 \hline
 \hline
 lens name & $\log_{10}(\sigma_{\delta\psi}^2)$ & $\beta$ & $\log_{10}(\sigma_{\delta\psi}^2)$ & $\beta$ & $\log_{10}(\sigma_{\delta\psi}^2)$ & $\beta$ & $\log_{10}(\sigma_{\delta\psi}^2)$ & $\beta$  \\ 
 \hline
J0029$-$0055 & $-2.85^{+0.45}_{-0.73}$ & $4.26^{+0.96}_{-0.76}$ & $-3.12^{+0.64}_{-0.82}$ & $4.52^{+1.36}_{-0.86}$ & $-3.09^{+0.61}_{-0.79}$ & $4.36^{+1.16}_{-0.76}$ & $-3.00^{+0.55}_{-0.82}$ & $4.41^{+1.16}_{-0.81}$ \\[1ex]
J0037$-$0942 & $-2.48^{+0.30}_{-0.36}$ & $4.97^{+0.96}_{-0.86}$ & $-2.73^{+0.45}_{-0.94}$ & $6.03^{+1.01}_{-1.31}$ & $-2.88^{+0.52}_{-0.94}$ & $5.63^{+1.21}_{-1.16}$ & $-2.64^{+0.39}_{-0.82}$ & $5.47^{+1.26}_{-1.11}$ \\[1ex]
J0252$+$0039 & $-3.09^{+0.52}_{-0.61}$ & $4.26^{+0.66}_{-0.71}$ & $-3.03^{+0.48}_{-0.64}$ & $4.26^{+0.71}_{-0.66}$ & $-3.12^{+0.55}_{-0.61}$ & $4.26^{+0.66}_{-0.71}$ & $-3.09^{+0.52}_{-0.61}$ & $4.26^{+0.71}_{-0.71}$ \\[1ex]
J0330$-$0020 & $-3.39^{+0.61}_{-0.61}$ & $4.11^{+0.66}_{-0.66}$ & $-3.52^{+0.61}_{-0.61}$ & $4.21^{+0.76}_{-0.71}$ & $-3.33^{+0.58}_{-0.61}$ & $4.11^{+0.71}_{-0.66}$ & $-3.39^{+0.58}_{-0.61}$ & $4.16^{+0.66}_{-0.71}$ \\[1ex]
J0728$+$3835 & $-2.79^{+0.45}_{-0.70}$ & $4.97^{+1.36}_{-1.01}$ & $-3.18^{+0.61}_{-0.79}$ & $4.87^{+1.57}_{-1.01}$ & $-3.15^{+0.61}_{-0.79}$ & $4.82^{+1.52}_{-1.01}$ & $-3.03^{+0.58}_{-0.79}$ & $4.87^{+1.52}_{-1.01}$ \\[1ex]
J0737$+$3216 & $-3.27^{+0.58}_{-0.64}$ & $4.31^{+0.71}_{-0.71}$ & $-3.55^{+0.64}_{-0.61}$ & $4.16^{+0.66}_{-0.66}$ & $-3.48^{+0.61}_{-0.58}$ & $4.11^{+0.71}_{-0.66}$ & $-3.42^{+0.61}_{-0.64}$ & $4.21^{+0.66}_{-0.71}$ \\[1ex]
J0903$+$4116 & $-2.73^{+0.42}_{-0.58}$ & $4.16^{+0.76}_{-0.66}$ & $-2.70^{+0.39}_{-0.64}$ & $4.31^{+1.01}_{-0.71}$ & $-2.88^{+0.48}_{-0.76}$ & $4.52^{+1.11}_{-0.81}$ & $-2.76^{+0.42}_{-0.67}$ & $4.31^{+0.96}_{-0.71}$ \\[1ex]
J0959$+$0410 & $-2.58^{+0.33}_{-0.45}$ & $3.96^{+0.71}_{-0.56}$ & $-3.27^{+0.70}_{-0.82}$ & $4.72^{+1.57}_{-0.96}$ & $-3.03^{+0.58}_{-0.82}$ & $4.72^{+1.31}_{-0.91}$ & $-2.88^{+0.52}_{-0.88}$ & $4.41^{+1.31}_{-0.81}$ \\[1ex]
J1112$+$0826 & $-2.61^{+0.36}_{-0.45}$ & $4.46^{+0.81}_{-0.71}$ & $-2.61^{+0.39}_{-0.58}$ & $4.97^{+1.11}_{-0.96}$ & $-2.55^{+0.33}_{-0.42}$ & $4.87^{+0.91}_{-0.76}$ & $-2.58^{+0.36}_{-0.48}$ & $4.77^{+0.96}_{-0.81}$ \\[1ex]
J1204$+$0358 & $-2.85^{+0.48}_{-0.73}$ & $5.12^{+0.96}_{-0.91}$ & $-3.39^{+0.76}_{-0.79}$ & $5.07^{+1.16}_{-0.96}$ & $-3.64^{+0.76}_{-0.67}$ & $4.97^{+1.26}_{-1.01}$ & $-3.27^{+0.70}_{-0.82}$ & $5.07^{+1.11}_{-0.96}$ \\[1ex]
J1250$+$0523 & $-2.61^{+0.33}_{-0.42}$ & $3.66^{+0.40}_{-0.40}$ & $-2.79^{+0.42}_{-0.58}$ & $3.96^{+0.66}_{-0.56}$ & $-2.73^{+0.39}_{-0.45}$ & $3.76^{+0.51}_{-0.45}$ & $-2.70^{+0.36}_{-0.48}$ & $3.76^{+0.56}_{-0.45}$ \\[1ex]
J1306$+$0600 & $-3.52^{+0.79}_{-0.67}$ & $4.52^{+1.06}_{-0.81}$ & $-3.30^{+0.70}_{-0.79}$ & $4.82^{+1.21}_{-0.91}$ & $-3.48^{+0.73}_{-0.73}$ & $4.72^{+1.16}_{-0.91}$ & $-3.45^{+0.76}_{-0.70}$ & $4.67^{+1.16}_{-0.86}$ \\[1ex]
J1313$+$4615 & $-2.76^{+0.42}_{-0.61}$ & $3.91^{+0.76}_{-0.56}$ & $-2.91^{+0.52}_{-0.94}$ & $4.67^{+1.72}_{-0.96}$ & $-3.00^{+0.55}_{-0.76}$ & $4.31^{+1.16}_{-0.76}$ & $-2.88^{+0.48}_{-0.79}$ & $4.26^{+1.31}_{-0.76}$ \\[1ex]
J1402$+$6321 & $-2.73^{+0.36}_{-0.45}$ & $3.96^{+0.56}_{-0.56}$ & $-2.85^{+0.45}_{-0.70}$ & $4.52^{+1.11}_{-0.81}$ & $-2.70^{+0.39}_{-0.45}$ & $4.16^{+0.66}_{-0.66}$ & $-2.76^{+0.39}_{-0.52}$ & $4.16^{+0.81}_{-0.66}$ \\[1ex]
J1531$-$0105 & $-3.21^{+0.61}_{-0.70}$ & $4.57^{+1.16}_{-0.86}$ & $-3.61^{+0.82}_{-0.70}$ & $5.53^{+1.21}_{-1.26}$ & $-3.45^{+0.73}_{-0.73}$ & $5.02^{+1.31}_{-1.06}$ & $-3.42^{+0.73}_{-0.73}$ & $4.97^{+1.41}_{-1.01}$ \\[1ex]
J1621$+$3931 & $-2.61^{+0.33}_{-0.33}$ & $3.56^{+0.40}_{-0.35}$ & $-2.58^{+0.33}_{-0.33}$ & $3.61^{+0.45}_{-0.40}$ & $-2.61^{+0.33}_{-0.39}$ & $3.66^{+0.40}_{-0.45}$ & $-2.58^{+0.30}_{-0.36}$ & $3.61^{+0.40}_{-0.40}$ \\[1ex]
J1627$-$0053 & $-3.09^{+0.45}_{-0.52}$ & $4.06^{+0.66}_{-0.61}$ & $-3.42^{+0.52}_{-0.55}$ & $4.01^{+0.61}_{-0.61}$ & $-3.45^{+0.52}_{-0.52}$ & $3.96^{+0.61}_{-0.61}$ & $-3.33^{+0.55}_{-0.52}$ & $4.01^{+0.61}_{-0.61}$ \\[1ex]
J1630$+$4520 & $-2.45^{+0.30}_{-0.30}$ & $5.12^{+0.86}_{-0.76}$ & $-2.70^{+0.39}_{-0.73}$ & $5.17^{+1.16}_{-0.96}$ & $-2.67^{+0.39}_{-0.70}$ & $5.42^{+1.01}_{-0.96}$ & $-2.58^{+0.36}_{-0.58}$ & $5.22^{+1.06}_{-0.86}$ \\[1ex]
J1636$+$4707 & $-2.85^{+0.45}_{-0.48}$ & $3.81^{+0.56}_{-0.51}$ & $-3.36^{+0.64}_{-0.70}$ & $4.36^{+1.01}_{-0.76}$ & $-3.09^{+0.52}_{-0.64}$ & $3.91^{+0.66}_{-0.56}$ & $-3.06^{+0.52}_{-0.67}$ & $4.01^{+0.71}_{-0.61}$ \\[1ex]
J2238$-$0754 & $-3.45^{+0.61}_{-0.61}$ & $4.16^{+0.66}_{-0.71}$ & $-3.94^{+0.52}_{-0.55}$ & $4.11^{+0.66}_{-0.66}$ & $-3.70^{+0.61}_{-0.61}$ & $4.11^{+0.71}_{-0.66}$ & $-3.73^{+0.64}_{-0.58}$ & $4.11^{+0.71}_{-0.66}$ \\[1ex]
J2300$+$0022 & $-3.61^{+0.79}_{-0.67}$ & $5.78^{+1.11}_{-1.26}$ & $-3.52^{+0.85}_{-0.76}$ & $6.23^{+0.96}_{-1.46}$ & $-3.70^{+0.70}_{-0.64}$ & $5.17^{+1.36}_{-1.11}$ & $-3.61^{+0.79}_{-0.7}$ & $5.73^{+1.21}_{-1.36}$ \\[1ex]
J2303$+$1422 & $-2.64^{+0.36}_{-0.48}$ & $4.36^{+0.76}_{-0.71}$ & $-3.15^{+0.61}_{-0.79}$ & $4.67^{+1.16}_{-0.91}$ & $-2.82^{+0.45}_{-0.67}$ & $4.36^{+0.91}_{-0.71}$ & $-2.82^{+0.45}_{-0.73}$ & $4.46^{+0.91}_{-0.76}$ \\[1ex]
J2343$-$0030 & $-2.85^{+0.42}_{-0.52}$ & $4.52^{+0.76}_{-0.71}$ & $-3.06^{+0.55}_{-0.73}$ & $4.82^{+0.96}_{-0.86}$ & $-2.97^{+0.48}_{-0.67}$ & $4.72^{+0.91}_{-0.81}$ & $-2.94^{+0.48}_{-0.64}$ & $4.67^{+0.86}_{-0.81}$ \\[1ex]
 \hline
\end{tabular}
\label{Table_measured_values}
\end{table*}

\section{Discussion and conclusions} \label{Conclusion_sec}

We demonstrated how ML can be used to predict the power-law statistics of a GRF perturbation to the lens potential by expanding on the framework of~\citet{Vernardos_2020} to make it more viable and reliable for observed data. In this study we used \emph{HST} images of 23 SLACS lenses, which will be similar to \textit{Euclid} in terms of data quality and resolution.
To apply ML to the tens of thousands of lenses expected to be found by \textit{Euclid} and the Vera C. Rubin observatory, a sufficiently diverse training set should be produced. For example, sources could be reconstructed by other ML methods such as~\citet{Biggio2022,Reconstructed_source1,Adam_2023} to build a training set that encapsulates all the known lenses. This could be combined with real galaxy images, simple analytic brightness profiles, or galaxy images from generative deep learning models~\citep[e.g.,][]{Lanusse_2021}. Ideally multiple training sets would be produced with different sources to ensure that the resulting predictions are independent of any particular assumption, like the training sets constructed in this work. A larger population of lenses will allow us to marginalize over the uncertainty in our predictions and more tightly constrain the joint probability distribution of the GRF power-law parameters, i.e. by narrowing the predictions in Fig.~\ref{real_data_average_prob_individual}. 

In this study, we assumed an SIE plus external shear smooth lens model and a GRF perturbation parameterized by a power-law, but our method could be used to quantify any type of perturbation on a smooth lens mass model. Consistent with~\citet{Chatterjee_2017,Vernardos_2020,Bayer_2023_1,Bayer_2023_2}, we parameterized the GRF power spectrum using a simple power-law (see equation~(\ref{power_law})), but a better parameterization could potentially be developed by analysing the power spectrum of numerical N-body and hydrodynamic simulations with different dark matter models. We also note that the GRF statistical framework works well at approximating the collective contribution of smaller mass dark matter substructure, but if the lens perturbation is dominated by very few large mass dark matter subhaloes, then this approximation breaks down~\citep{Vegetti_2010,Hezaveh_2016,Chatterjee_2017}. In future work, the low mass dark matter subhaloes could be modelled as a GRF but include possible contamination by the occasional large mass subhalo that can have a substantial localized effect on the lens potential. Improvements could also be made to disentangle baryonic substructure from the dark matter substructure with the inclusion of boxyness, discyness, or other baryonic processes in the smooth mass profile~\citep[see,][]{Vyvere_2022}. 

Our ML model may also be improved in future work. Following~\citet{Vernardos_2020}, our method predicts marginal probability distributions for our power-law parameters which makes our predicted uncertainties uncorrelated. Our method could be further developed to include correlations between the substructure parameters in the uncertainty. This would be especially important if the GRF was parameterized by a more complicated power spectrum where the parameters could include significant degeneracies. The use of vision transformers could also be considered, as they may show improved performance compared to CNNs in strong lens parameter estimation~\citep{huang2022strong}.

While we showed that our reported uncertainty range is appropriate across the test set of mock images, there may be additional systematic uncertainties when applying the CNN to real lenses which we cannot fully account for. We partially addressed this by training three separate models using different source methods, which~\citet{Vernardos_2020} found to be the most important consideration. The sources from~\citet{Shajib_2021} may deviate from the true sources of our lenses. In data set 2 we used the entire MCMC chain, so any source discrepancies would have to be completely outside the posterior distribution of the MCMC chain to bias our network. We also compare the results of using the network trained on the entire MCMC chain to the network trained with sources produced through data augmentation of the best fitting sources and find agreement in the results. The data augmentation procedure can lead to large deviations in the source morphology, so the true sources would have to be completely outside that posterior distribution as well. Other choices in building the training set could also affect our results. For example, when simulating our mock lenses we assumed a Sérsic lens light profile, while the real data may have more complicated lens light profiles which often require 2 Sérsic profiles to fit. Discrepancies in uncertainty estimations may always exist when a neural network is trained and tested with simulated data and then applied to real data. We attempted to bridge some of these discrepancies by using separately trained CNNs with complementary training sets containing different sources and then averaging the resulting predicted probability distributions, increasing the reported uncertainty. Further diversification could also be used by altering the smooth lens model, the lens light model, or the observational effects across training sets. 

Thus far only a single lens has had its GRF power spectrum constrained outside of this work by analysing the power spectrum of surface brightness anomalies in~\citet{Bayer_2023_2}. They constrained the GRF power spectrum in lens J0252$+$0039 to have $\log_{10}(\sigma_{\delta\psi}^2) < -2.5$ at the 99 per cent confidence interval for $3 \leq\beta\leq 8$. This is consistent with our predictions where we find average values of $\log_{10}(\sigma_{\delta\psi}^2)= -3.09^{+0.52}_{-0.61}$ and $\beta = 4.26^{+0.71}_{-0.71}$. Follow up studies are required to determine the consistency between our two approaches on a larger lens sample. 

For future observations, a combination of ML approaches like ours and more traditional lens modelling approaches like in~\citet{Bayer_2023_2} may give better results than using either method individually. A pretrained neural network could be used to first give an estimate of the GRF perturbation parameters across the tens of thousands of lenses expected to be observed by next generation surveys in just a few minutes of computational time with a single GPU. Lenses with interesting properties could be selected for follow up studies using a computationally expensive lens modelling approach to further constrain the lens properties and double-check the ML predictions. These methods can take weeks of computational time and often need to be fine tuned to each system, making them unrealistic to apply to all the lenses. As of now, the method of~\citep{Bayer_2023_2} only produces upper bound constraints on the GRF parameters while our ML method can make direct predictions. In future work, our predictions could also be compared with the predictions from hydrodynamic simulations such as EAGLE~\citep{Cosmology3} and IllustrisTNG~\citep{nelson2021illustristng}.

\section*{Acknowledgements}

This research was made possible by the generosity of Eric and Wendy Schmidt by recommendation of the Schmidt Futures program. G.V. has received funding from the European Union's Horizon 2020 research and innovation programme under the Marie Sklodovska-Curie grant agreement No 897124. G.T. was funded by the CALCHAS project (contract no. 842560) within the H2020 Framework Program of the European Commission. The data used in this publication were collected through the MENDEL high performance computing (HPC) cluster at the American Museum of Natural History. This HPC cluster was developed with National Science Foundation (NSF) Campus Cyberinfrastructure support through Award \#1925590.

This work utilised several open source libraries including: \texttt{MOLET}~\citep{MOLET}, 
\texttt{TensorFlow}~\citep{tensorflow2015}
\texttt{Matplotlib}~\citep{Matplotlib}, \texttt{Numpy}~\citep{Numpy}, \texttt{Scipy}~\citep{Scipy}, 
\texttt{lenstronomy}~\citep{lenstronomy_paper1,lenstronomy_paper2},
\texttt{Astropy}~\citep{Astropy}, and \texttt{corner.py}~\citep{corner}.

\section*{Data availability}
The \emph{HST} imaging data used in this paper are publicly available from the Mikulski Archive for Space Telescopes (MAST). The simulated data utilised in this paper are available upon reasonable request from the corresponding author. The machine learning code and predicted probability distributions for our lens sample are publicly available at:~\url{https://github.com/JFagin/GRF_ML}. 

\bibliographystyle{mnras}
\bibliography{bib.bib}

\begin{thebibliography}{}
\makeatletter
\relax
\def\mn@urlcharsother{\let\do\@makeother \do\$\do\&\do\#\do\^\do\_\do\%\do\~}
\def\mn@doi{\begingroup\mn@urlcharsother \@ifnextchar [ {\mn@doi@}
  {\mn@doi@[]}}
\def\mn@doi@[#1]#2{\def\@tempa{#1}\ifx\@tempa\@empty \href
  {http://dx.doi.org/#2} {doi:#2}\else \href {http://dx.doi.org/#2} {#1}\fi
  \endgroup}
\def\mn@eprint#1#2{\mn@eprint@#1:#2::\@nil}
\def\mn@eprint@arXiv#1{\href {http://arxiv.org/abs/#1} {{\tt arXiv:#1}}}
\def\mn@eprint@dblp#1{\href {http://dblp.uni-trier.de/rec/bibtex/#1.xml}
  {dblp:#1}}
\def\mn@eprint@#1:#2:#3:#4\@nil{\def\@tempa {#1}\def\@tempb {#2}\def\@tempc
  {#3}\ifx \@tempc \@empty \let \@tempc \@tempb \let \@tempb \@tempa \fi \ifx
  \@tempb \@empty \def\@tempb {arXiv}\fi \@ifundefined
  {mn@eprint@\@tempb}{\@tempb:\@tempc}{\expandafter \expandafter \csname
  mn@eprint@\@tempb\endcsname \expandafter{\@tempc}}}

\bibitem[\protect\citeauthoryear{Abadi et~al.,}{Abadi
  et~al.}{2015}]{tensorflow2015}
Abadi M.,  et~al., 2015, {TensorFlow}: Large-Scale Machine Learning on
  Heterogeneous Systems, \url {https://www.tensorflow.org/}

\bibitem[\protect\citeauthoryear{Adam, Perreault-Levasseur, Hezaveh  \&
  Welling}{Adam et~al.}{2023}]{Adam_2023}
Adam A.,  Perreault-Levasseur L.,  Hezaveh Y.,   Welling M.,  2023, \mn@doi
  [The Astrophysical Journal] {10.3847/1538-4357/accf84}, 951, 6

\bibitem[\protect\citeauthoryear{Aghanim et~al.,}{Aghanim
  et~al.}{2020}]{Cosmology2}
Aghanim N.,  et~al., 2020, \mn@doi [Astronomy & Astrophysics]
  {10.1051/0004-6361/201833910}, 641, A6

\bibitem[\protect\citeauthoryear{Alexander, Gleyzer, McDonough, Toomey  \&
  Usai}{Alexander et~al.}{2020}]{Dark_matter_CNN_3}
Alexander S.,  Gleyzer S.,  McDonough E.,  Toomey M.~W.,   Usai E.,  2020,
  \mn@doi [The Astrophysical Journal] {10.3847/1538-4357/ab7925}, 893, 15

\bibitem[\protect\citeauthoryear{Alexander, Gleyzer, Reddy, Tidball  \&
  Toomey}{Alexander et~al.}{2021}]{Domain_adaptation}
Alexander S.,  Gleyzer S.,  Reddy P.,  Tidball M.,   Toomey M.~W.,  2021,
  Domain Adaptation for Simulation-Based Dark Matter Searches Using Strong
  Gravitational Lensing (\mn@eprint {arXiv} {2112.12121})

\bibitem[\protect\citeauthoryear{{Astropy Collaboration} et~al.,}{{Astropy
  Collaboration} et~al.}{2018}]{Astropy}
{Astropy Collaboration} et~al., 2018, \mn@doi [\aj] {10.3847/1538-3881/aabc4f},
  \href {https://ui.adsabs.harvard.edu/abs/2018AJ....156..123A} {156, 123}

\bibitem[\protect\citeauthoryear{Auger, Treu, Bolton, Gavazzi, Koopmans,
  Marshall, Bundy  \& Moustakas}{Auger et~al.}{2009}]{Sloan_ACS_lenses}
Auger M.~W.,  Treu T.,  Bolton A.~S.,  Gavazzi R.,  Koopmans L. V.~E.,
  Marshall P.~J.,  Bundy K.,   Moustakas L.~A.,  2009, \mn@doi [The
  Astrophysical Journal] {10.1088/0004-637x/705/2/1099}, 705, 1099–1115

\bibitem[\protect\citeauthoryear{Avila, Hack, Cara, Borncamp, Mack, Smith  \&
  Ubeda}{Avila et~al.}{2014}]{Astro_drizzel_package}
Avila R.~J.,  Hack W.,  Cara M.,  Borncamp D.,  Mack J.,  Smith L.,   Ubeda L.,
   2014, DrizzlePac 2.0 - Introducing New Features (\mn@eprint {arXiv}
  {1411.5605})

\bibitem[\protect\citeauthoryear{{Baggett}, {Gosmeyer}  \& {Noeske}}{{Baggett}
  et~al.}{2015}]{CTI1}
{Baggett} S.,  {Gosmeyer} C.,   {Noeske} K.,  2015, {WFC3/UVIS Charge Transfer
  Efficiency 2009-2015}, Space Telescope WFC Instrument Science Report

\bibitem[\protect\citeauthoryear{Bayer, Koopmans, McKean, Vegetti, Treu,
  Fassnacht  \& Glazebrook}{Bayer et~al.}{2023a}]{Bayer_2023_1}
Bayer D.,  Koopmans L. V.~E.,  McKean J.~P.,  Vegetti S.,  Treu T.,  Fassnacht
  C.~D.,   Glazebrook K.,  2023a, \mn@doi [Monthly Notices of the Royal
  Astronomical Society] {10.1093/mnras/stad1403}, 523, 1326

\bibitem[\protect\citeauthoryear{Bayer, Chatterjee, Koopmans, Vegetti, McKean,
  Treu, Fassnacht  \& Glazebrook}{Bayer et~al.}{2023b}]{Bayer_2023_2}
Bayer D.,  Chatterjee S.,  Koopmans L. V.~E.,  Vegetti S.,  McKean J.~P.,  Treu
  T.,  Fassnacht C.~D.,   Glazebrook K.,  2023b, \mn@doi [Monthly Notices of
  the Royal Astronomical Society] {10.1093/mnras/stad1402}, 523, 1310

\bibitem[\protect\citeauthoryear{Biggio, Vernardos, Galan  \& Peel}{Biggio
  et~al.}{2022}]{Biggio2022}
Biggio L.,  Vernardos G.,  Galan A.,   Peel A.,  2022, Modeling lens potentials
  with continuous neural fields in galaxy-scale strong lenses (\mn@eprint
  {arXiv} {2210.09169})

\bibitem[\protect\citeauthoryear{Birrer \& Amara}{Birrer \&
  Amara}{2018}]{lenstronomy_paper1}
Birrer S.,  Amara A.,  2018, Lenstronomy: multi-purpose gravitational lens
  modelling software package (\mn@eprint {arXiv} {1803.09746})

\bibitem[\protect\citeauthoryear{Birrer, Amara  \& Refregier}{Birrer
  et~al.}{2015}]{Lens_modeling_with_basis}
Birrer S.,  Amara A.,   Refregier A.,  2015, \mn@doi [The Astrophysical
  Journal] {10.1088/0004-637x/813/2/102}, 813, 102

\bibitem[\protect\citeauthoryear{Birrer et~al.,}{Birrer
  et~al.}{2019}]{Used_Shapelets_Hubble1}
Birrer S.,  et~al., 2019, \mn@doi [Monthly Notices of the Royal Astronomical
  Society] {10.1093/mnras/stz200}, 484, 4726–4753

\bibitem[\protect\citeauthoryear{Birrer et~al.,}{Birrer
  et~al.}{2021}]{lenstronomy_paper2}
Birrer S.,  et~al., 2021, \mn@doi [Journal of Open Source Software]
  {10.21105/joss.03283}, 6, 3283

\bibitem[\protect\citeauthoryear{Bullock \& Boylan-Kolchin}{Bullock \&
  Boylan-Kolchin}{2017}]{Dark_matter_on_small_scale}
Bullock J.~S.,  Boylan-Kolchin M.,  2017, \mn@doi [Annual Review of Astronomy
  and Astrophysics] {10.1146/annurev-astro-091916-055313}, 55, 343–387

\bibitem[\protect\citeauthoryear{Casertano et~al.,}{Casertano
  et~al.}{2000}]{Drizzle_effect_Hubble}
Casertano S.,  et~al., 2000, \mn@doi [The Astronomical Journal]
  {10.1086/316851}, 120, 2747–2824

\bibitem[\protect\citeauthoryear{Chatterjee \& Koopmans}{Chatterjee \&
  Koopmans}{2017}]{Chatterjee_2017}
Chatterjee S.,  Koopmans L. V.~E.,  2017, \mn@doi [Monthly Notices of the Royal
  Astronomical Society] {10.1093/mnras/stx2674}, 474, 1762–1772

\bibitem[\protect\citeauthoryear{Collett}{Collett}{2015}]{Number_Strong_Lens_Discovered}
Collett T.,  2015, \mn@doi [The Astrophysical Journal]
  {10.1088/0004-637X/811/1/20}, 811

\bibitem[\protect\citeauthoryear{Diaz~Rivero \& Dvorkin}{Diaz~Rivero \&
  Dvorkin}{2020}]{Dark_matter_CNN_2}
Diaz~Rivero A.,  Dvorkin C.,  2020, \mn@doi [Physical Review D]
  {10.1103/physrevd.101.023515}, 101

\bibitem[\protect\citeauthoryear{Foreman-Mackey}{Foreman-Mackey}{2016}]{corner}
Foreman-Mackey D.,  2016, \mn@doi [The Journal of Open Source Software]
  {10.21105/joss.00024}, 1, 24

\bibitem[\protect\citeauthoryear{Fruchter \& Hook}{Fruchter \&
  Hook}{2002}]{Drizzling_correlated_noise}
Fruchter A.,  Hook R.,  2002, \mn@doi [Publications of the Astronomical Society
  of the Pacific] {10.1086/338393}, 114, 144–152

\bibitem[\protect\citeauthoryear{Fu, Li, Liu, Gao, Celikyilmaz  \& Carin}{Fu
  et~al.}{2019}]{KL_annealing}
Fu H.,  Li C.,  Liu X.,  Gao J.,  Celikyilmaz A.,   Carin L.,  2019, Cyclical
  Annealing Schedule: A Simple Approach to Mitigating KL Vanishing,
  \mn@doi{10.48550/ARXIV.1903.10145}, \url {https://arxiv.org/abs/1903.10145}

\bibitem[\protect\citeauthoryear{Gal \& Ghahramani}{Gal \&
  Ghahramani}{2016}]{MC_Dropout}
Gal Y.,  Ghahramani Z.,  2016, Dropout as a Bayesian Approximation:
  Representing Model Uncertainty in Deep Learning (\mn@eprint {arXiv}
  {1506.02142})

\bibitem[\protect\citeauthoryear{Galan, Peel, Joseph, Courbin  \& Starck}{Galan
  et~al.}{2021}]{galan2021}
Galan A.,  Peel A.,  Joseph R.,  Courbin F.,   Starck J.-L.,  2021, \mn@doi
  [Astronomy \& Astrophysics] {10.1051/0004-6361/202039363}, 647, A176

\bibitem[\protect\citeauthoryear{Galan, Vernardos, Peel, Courbin  \&
  Starck}{Galan et~al.}{2022}]{Galan2022}
Galan A.,  Vernardos G.,  Peel A.,  Courbin F.,   Starck J.-L.,  2022, \mn@doi
  [Astronomy \& Astrophysics] {10.1051/0004-6361/202244464}, 668, A155

\bibitem[\protect\citeauthoryear{Ganaie, Hu, Malik, Tanveer  \&
  Suganthan}{Ganaie et~al.}{2022}]{Ganaie_2022}
Ganaie M.,  Hu M.,  Malik A.,  Tanveer M.,   Suganthan P.,  2022, \mn@doi
  [Engineering Applications of Artificial Intelligence]
  {10.1016/j.engappai.2022.105151}, 115, 105151

\bibitem[\protect\citeauthoryear{Gavazzi, Treu, Rhodes, Koopmans, Bolton,
  Burles, Massey  \& Moustakas}{Gavazzi et~al.}{2007}]{Isothermal1}
Gavazzi R.,  Treu T.,  Rhodes J.~D.,  Koopmans L. V.~E.,  Bolton A.~S.,  Burles
  S.,  Massey R.~J.,   Moustakas L.~A.,  2007, \mn@doi [The Astrophysical
  Journal] {10.1086/519237}, 667, 176–190

\bibitem[\protect\citeauthoryear{Graves}{Graves}{2011}]{VI}
Graves A.,  2011, in Proceedings of the 24th International Conference on Neural
  Information Processing Systems. NIPS'11.
Curran Associates Inc., Red Hook, NY, USA, p. 2348–2356

\bibitem[\protect\citeauthoryear{Harris et~al.,}{Harris et~al.}{2020}]{Numpy}
Harris C.~R.,  et~al., 2020, \mn@doi [Nature] {10.1038/s41586-020-2649-2}, 585,
  357–362

\bibitem[\protect\citeauthoryear{He, Zhang, Ren  \& Sun}{He
  et~al.}{2015}]{Resnet2015}
He K.,  Zhang X.,  Ren S.,   Sun J.,  2015, Deep Residual Learning for Image
  Recognition (\mn@eprint {arXiv} {1512.03385})

\bibitem[\protect\citeauthoryear{Hezaveh et~al.,}{Hezaveh
  et~al.}{2016a}]{Subhalo_mass3}
Hezaveh Y.~D.,  et~al., 2016a, \mn@doi [The Astrophysical Journal]
  {10.3847/0004-637x/823/1/37}, 823, 37

\bibitem[\protect\citeauthoryear{Hezaveh, Dalal, Holder, Kisner, Kuhlen  \&
  Levasseur}{Hezaveh et~al.}{2016b}]{Hezaveh_2016}
Hezaveh Y.,  Dalal N.,  Holder G.,  Kisner T.,  Kuhlen M.,   Levasseur L.~P.,
  2016b, \mn@doi [Journal of Cosmology and Astroparticle Physics]
  {10.1088/1475-7516/2016/11/048}, 2016, 048

\bibitem[\protect\citeauthoryear{Hezaveh, Levasseur  \& Marshall}{Hezaveh
  et~al.}{2017}]{CNN_without_uncertainties}
Hezaveh Y.~D.,  Levasseur L.~P.,   Marshall P.~J.,  2017, \mn@doi [Nature]
  {10.1038/nature23463}, 548, 555–557

\bibitem[\protect\citeauthoryear{Hort{\'{u}}a, Volpi, Marinelli  \&
  Malag{\`{o}}}{Hort{\'{u}}a et~al.}{2020}]{CMB_Flipout1}
Hort{\'{u}}a H.~J.,  Volpi R.,  Marinelli D.,   Malag{\`{o}} L.,  2020, \mn@doi
  [Physical Review D] {10.1103/physrevd.102.103509}, 102

\bibitem[\protect\citeauthoryear{Hu, Shen, Albanie, Sun  \& Wu}{Hu
  et~al.}{2017}]{Squeeze_and_excitation}
Hu J.,  Shen L.,  Albanie S.,  Sun G.,   Wu E.,  2017, Squeeze-and-Excitation
  Networks, \mn@doi{10.48550/ARXIV.1709.01507}, \url
  {https://arxiv.org/abs/1709.01507}

\bibitem[\protect\citeauthoryear{Huang, Chen, Chang, Lin, Hsu, Thengane  \&
  Lin}{Huang et~al.}{2022}]{huang2022strong}
Huang K.-W.,  Chen G. C.-F.,  Chang P.-W.,  Lin S.-C.,  Hsu C.-J.,  Thengane
  V.,   Lin J. Y.-Y.,  2022, Strong Gravitational Lensing Parameter Estimation
  with Vision Transformer (\mn@eprint {arXiv} {2210.04143})

\bibitem[\protect\citeauthoryear{Hunter}{Hunter}{2007}]{Matplotlib}
Hunter J.~D.,  2007, \mn@doi [Computing in Science \& Engineering]
  {10.1109/MCSE.2007.55}, 9, 90

\bibitem[\protect\citeauthoryear{Ivezić et~al.,}{Ivezić et~al.}{2019}]{LSST}
Ivezić c.,  et~al., 2019, \mn@doi [The Astrophysical Journal]
  {10.3847/1538-4357/ab042c}, 873, 111

\bibitem[\protect\citeauthoryear{{Jacobs}, {Glazebrook}, {Collett}, {More}  \&
  {McCarthy}}{{Jacobs} et~al.}{2017}]{Strong_Lensing_Detection_With_CNN_1}
{Jacobs} C.,  {Glazebrook} K.,  {Collett} T.,  {More} A.,   {McCarthy} C.,
  2017, \mn@doi [\mnras] {10.1093/mnras/stx1492}, \href
  {https://ui.adsabs.harvard.edu/abs/2017MNRAS.471..167J} {471, 167}

\bibitem[\protect\citeauthoryear{Jullo, Kneib, Limousin, Elíasdóttir,
  Marshall  \& Verdugo}{Jullo et~al.}{2007}]{MCMC_lens_modeling}
Jullo E.,  Kneib J.-P.,  Limousin M.,  Elíasdóttir A.,  Marshall P.~J.,
  Verdugo T.,  2007, \mn@doi [New Journal of Physics]
  {10.1088/1367-2630/9/12/447}, 9, 447–447

\bibitem[\protect\citeauthoryear{Karchev, Coogan  \& Weniger}{Karchev
  et~al.}{2022}]{Reconstructed_source1}
Karchev K.,  Coogan A.,   Weniger C.,  2022, \mn@doi [Monthly Notices of the
  Royal Astronomical Society] {10.1093/mnras/stac311}, 512, 661

\bibitem[\protect\citeauthoryear{{Kassiola} \& {Kovner}}{{Kassiola} \&
  {Kovner}}{1993}]{SIE_1993}
{Kassiola} A.,  {Kovner} I.,  1993, \mn@doi [\apj] {10.1086/173325}, \href
  {https://ui.adsabs.harvard.edu/abs/1993ApJ...417..450K} {417, 450}

\bibitem[\protect\citeauthoryear{{Keeton} \& {Kochanek}}{{Keeton} \&
  {Kochanek}}{1998}]{SIE_1998}
{Keeton} C.~R.,  {Kochanek} C.~S.,  1998, \mn@doi [\apj] {10.1086/305272},
  \href {https://ui.adsabs.harvard.edu/abs/1998ApJ...495..157K} {495, 157}

\bibitem[\protect\citeauthoryear{Kendall \& Gal}{Kendall \&
  Gal}{2017}]{kendall2017uncertainties}
Kendall A.,  Gal Y.,  2017, What Uncertainties Do We Need in Bayesian Deep
  Learning for Computer Vision? (\mn@eprint {arXiv} {1703.04977})

\bibitem[\protect\citeauthoryear{Kingma \& Ba}{Kingma \& Ba}{2017}]{Adam}
Kingma D.~P.,  Ba J.,  2017, Adam: A Method for Stochastic Optimization
  (\mn@eprint {arXiv} {1412.6980})

\bibitem[\protect\citeauthoryear{Komatsu et~al.,}{Komatsu
  et~al.}{2011}]{Cosmology1}
Komatsu E.,  et~al., 2011, \mn@doi [The Astrophysical Journal Supplement
  Series] {10.1088/0067-0049/192/2/18}, 192, 18

\bibitem[\protect\citeauthoryear{Koopmans}{Koopmans}{2005}]{Koopmans_2005}
Koopmans L. V.~E.,  2005, \mn@doi [Monthly Notices of the Royal Astronomical
  Society] {10.1111/j.1365-2966.2005.09523.x}, 363, 1136

\bibitem[\protect\citeauthoryear{{Kormann}, {Schneider}  \&
  {Bartelmann}}{{Kormann} et~al.}{1994}]{SIE_1994}
{Kormann} R.,  {Schneider} P.,   {Bartelmann} M.,  1994, \aap, \href
  {https://ui.adsabs.harvard.edu/abs/1994A&A...284..285K} {284, 285}

\bibitem[\protect\citeauthoryear{{Krist}, {Hook}  \& {Stoehr}}{{Krist}
  et~al.}{2011}]{Tiny_Tim}
{Krist} J.~E.,  {Hook} R.~N.,   {Stoehr} F.,  2011, in {Kahan} M.~A.,  ed.,
  Society of Photo-Optical Instrumentation Engineers (SPIE) Conference Series
  Vol. 8127, Optical Modeling and Performance Predictions V. p. 81270J,
  \mn@doi{10.1117/12.892762}

\bibitem[\protect\citeauthoryear{Lanusse, Mandelbaum, Ravanbakhsh, Li, Freeman
  \& Póczos}{Lanusse et~al.}{2021}]{Lanusse_2021}
Lanusse F.,  Mandelbaum R.,  Ravanbakhsh S.,  Li C.-L.,  Freeman P.,   Póczos
  B.,  2021, \mn@doi [Monthly Notices of the Royal Astronomical Society]
  {10.1093/mnras/stab1214}, 504, 5543–5555

\bibitem[\protect\citeauthoryear{Laureijs et~al.,}{Laureijs
  et~al.}{2011}]{Euclid}
Laureijs R.,  et~al., 2011, Euclid Definition Study Report (\mn@eprint {arXiv}
  {1110.3193})

\bibitem[\protect\citeauthoryear{Li, Frenk, Cole, Gao, Bose  \& Hellwing}{Li
  et~al.}{2016}]{Subhalo_mass2}
Li R.,  Frenk C.~S.,  Cole S.,  Gao L.,  Bose S.,   Hellwing W.~A.,  2016,
  \mn@doi [Monthly Notices of the Royal Astronomical Society]
  {10.1093/mnras/stw939}, 460, 363

\bibitem[\protect\citeauthoryear{Massey, Stoughton, Leauthaud, Rhodes,
  Koekemoer, Ellis  \& Shaghoulian}{Massey et~al.}{2009}]{CTI2}
Massey R.,  Stoughton C.,  Leauthaud A.,  Rhodes J.,  Koekemoer A.,  Ellis R.,
   Shaghoulian E.,  2009, \mn@doi [Monthly Notices of the Royal Astronomical
  Society] {10.1111/j.1365-2966.2009.15638.x}, 401, 371

\bibitem[\protect\citeauthoryear{Metcalf et~al.,}{Metcalf
  et~al.}{2019}]{Strong_Lensing_Detection_With_CNN_3}
Metcalf R.~B.,  et~al., 2019, \mn@doi [Astronomy & Astrophysics]
  {10.1051/0004-6361/201832797}, 625, A119

\bibitem[\protect\citeauthoryear{Montel, Coogan, Correa, Karchev  \&
  Weniger}{Montel et~al.}{2022}]{CNN_Substructure_Mass_Function2}
Montel N.~A.,  Coogan A.,  Correa C.,  Karchev K.,   Weniger C.,  2022,
  Estimating the warm dark matter mass from strong lensing images with
  truncated marginal neural ratio estimation,
  \mn@doi{10.48550/ARXIV.2205.09126}, \url {https://arxiv.org/abs/2205.09126}

\bibitem[\protect\citeauthoryear{Navarro, Frenk  \& White}{Navarro
  et~al.}{1997}]{NFW}
Navarro J.~F.,  Frenk C.~S.,   White S. D.~M.,  1997, \mn@doi [The
  Astrophysical Journal] {10.1086/304888}, 490, 493–508

\bibitem[\protect\citeauthoryear{Nelson et~al.,}{Nelson
  et~al.}{2021}]{nelson2021illustristng}
Nelson D.,  et~al., 2021, The IllustrisTNG Simulations: Public Data Release
  (\mn@eprint {arXiv} {1812.05609})

\bibitem[\protect\citeauthoryear{Nightingale et~al.,}{Nightingale
  et~al.}{2021}]{Nightingale_2021}
Nightingale J.,  et~al., 2021, \mn@doi [Journal of Open Source Software]
  {10.21105/joss.02825}, 6, 2825

\bibitem[\protect\citeauthoryear{Oldham \& Auger}{Oldham \&
  Auger}{2018}]{Isothermal4}
Oldham L.~J.,  Auger M.~W.,  2018, \mn@doi [Monthly Notices of the Royal
  Astronomical Society] {10.1093/mnras/sty065}, 476, 133

\bibitem[\protect\citeauthoryear{Ostdiek, Rivero  \& Dvorkin}{Ostdiek
  et~al.}{2021}]{Subhalo_mass_using_machine_learning}
Ostdiek B.,  Rivero A.~D.,   Dvorkin C.,  2021, Extracting the Subhalo Mass
  Function from Strong Lens Images with Image Segmentation (\mn@eprint {arXiv}
  {2009.06639})

\bibitem[\protect\citeauthoryear{Pearson, Li  \& Dye}{Pearson
  et~al.}{2019}]{Estimating_SIE_Par_Using_CNN}
Pearson J.,  Li N.,   Dye S.,  2019, \mn@doi [Monthly Notices of the Royal
  Astronomical Society] {10.1093/mnras/stz1750}, 488, 991–1004

\bibitem[\protect\citeauthoryear{Pearson, Maresca, Li  \& Dye}{Pearson
  et~al.}{2021}]{Peason2021}
Pearson J.,  Maresca J.,  Li N.,   Dye S.,  2021, \mn@doi [Monthly Notices of
  the Royal Astronomical Society] {10.1093/mnras/stab1547}, 505, 4362–4382

\bibitem[\protect\citeauthoryear{Perreault~Levasseur, Hezaveh  \&
  Wechsler}{Perreault~Levasseur et~al.}{2017}]{Uncertainty_in_SIE_predictions}
Perreault~Levasseur L.,  Hezaveh Y.~D.,   Wechsler R.~H.,  2017, \mn@doi [The
  Astrophysical Journal] {10.3847/2041-8213/aa9704}, 850, L7

\bibitem[\protect\citeauthoryear{Petrillo et~al.,}{Petrillo
  et~al.}{2017}]{Strong_Lensing_Detection_With_CNN_2}
Petrillo C.~E.,  et~al., 2017, \mn@doi [Monthly Notices of the Royal
  Astronomical Society] {10.1093/mnras/stx2052}, 472, 1129–1150

\bibitem[\protect\citeauthoryear{Refregier}{Refregier}{2003}]{Shapelets}
Refregier A.,  2003, \mn@doi [Monthly Notices of the Royal Astronomical
  Society] {10.1046/j.1365-8711.2003.05901.x}, 338, 35–47

\bibitem[\protect\citeauthoryear{Rojas et~al.,}{Rojas
  et~al.}{2022}]{Rojas_2022}
Rojas K.,  et~al., 2022, \mn@doi [Astronomy &amp; Astrophysics]
  {10.1051/0004-6361/202142119}, 668, A73

\bibitem[\protect\citeauthoryear{Savary et~al.,}{Savary
  et~al.}{2022}]{Savary2022}
Savary E.,  et~al., 2022, \mn@doi [Astronomy \& Astrophysics]
  {10.1051/0004-6361/202142505}, 666, A1

\bibitem[\protect\citeauthoryear{{Schaefer}, {Geiger}, {Kuntzer}  \&
  {Kneib}}{{Schaefer} et~al.}{2018}]{Strong_Lensing_Detection_With_CNN_4}
{Schaefer} C.,  {Geiger} M.,  {Kuntzer} T.,   {Kneib} J.~P.,  2018, \mn@doi
  [\aap] {10.1051/0004-6361/201731201}, \href
  {https://ui.adsabs.harvard.edu/abs/2018A&A...611A...2S} {611, A2}

\bibitem[\protect\citeauthoryear{Schaye et~al.,}{Schaye
  et~al.}{2014}]{Cosmology3}
Schaye J.,  et~al., 2014, \mn@doi [Monthly Notices of the Royal Astronomical
  Society] {10.1093/mnras/stu2058}, 446, 521–554

\bibitem[\protect\citeauthoryear{{Schuldt}, {Suyu}, {Meinhardt},
  {Leal-Taix{\'e}}, {Ca{\~n}ameras}, {Taubenberger}  \& {Halkola}}{{Schuldt}
  et~al.}{2021}]{Schuldt2021}
{Schuldt} S.,  {Suyu} S.~H.,  {Meinhardt} T.,  {Leal-Taix{\'e}} L.,
  {Ca{\~n}ameras} R.,  {Taubenberger} S.,   {Halkola} A.,  2021, \mn@doi [\aap]
  {10.1051/0004-6361/202039574}, \href
  {https://ui.adsabs.harvard.edu/abs/2021A&A...646A.126S} {646, A126}

\bibitem[\protect\citeauthoryear{{Sersic}}{{Sersic}}{1968}]{Original_Sersic_Paper}
{Sersic} J.~L.,  1968, {Atlas de Galaxias Australes}

\bibitem[\protect\citeauthoryear{{Shajib} et~al.,}{{Shajib}
  et~al.}{2019}]{Used_Shapelets_Quasar}
{Shajib} A.~J.,  et~al., 2019, \mn@doi [\mnras] {10.1093/mnras/sty3397}, \href
  {https://ui.adsabs.harvard.edu/abs/2019MNRAS.483.5649S} {483, 5649}

\bibitem[\protect\citeauthoryear{Shajib et~al.,}{Shajib
  et~al.}{2020}]{Used_Shapelets_Hubble2}
Shajib A.~J.,  et~al., 2020, \mn@doi [Monthly Notices of the Royal Astronomical
  Society] {10.1093/mnras/staa828}, 494, 6072–6102

\bibitem[\protect\citeauthoryear{Shajib, Treu, Birrer  \& Sonnenfeld}{Shajib
  et~al.}{2021}]{Shajib_2021}
Shajib A.~J.,  Treu T.,  Birrer S.,   Sonnenfeld A.,  2021, \mn@doi [Monthly
  Notices of the Royal Astronomical Society] {10.1093/mnras/stab536}, 503,
  2380–2405

\bibitem[\protect\citeauthoryear{Shajib et~al.,}{Shajib
  et~al.}{2022}]{shajib2022strong}
Shajib A.~J.,  et~al., 2022, Strong Lensing by Galaxies (\mn@eprint {arXiv}
  {2210.10790})

\bibitem[\protect\citeauthoryear{{Suyu} et~al.,}{{Suyu}
  et~al.}{2014}]{Isothermal3}
{Suyu} S.~H.,  et~al., 2014, \mn@doi [\apjl] {10.1088/2041-8205/788/2/L35},
  \href {https://ui.adsabs.harvard.edu/abs/2014ApJ...788L..35S} {788, L35}

\bibitem[\protect\citeauthoryear{Tagore \& Jackson}{Tagore \&
  Jackson}{2016}]{Shapelets_for_lense_modeling}
Tagore A.~S.,  Jackson N.,  2016, \mn@doi [Monthly Notices of the Royal
  Astronomical Society] {10.1093/mnras/stw057}, 457, 3066–3075

\bibitem[\protect\citeauthoryear{{Treu}, {Auger}, {Koopmans}, {Gavazzi},
  {Marshall}  \& {Bolton}}{{Treu} et~al.}{2010}]{Isothermal2}
{Treu} T.,  {Auger} M.~W.,  {Koopmans} L. V.~E.,  {Gavazzi} R.,  {Marshall}
  P.~J.,   {Bolton} A.~S.,  2010, \mn@doi [\apj]
  {10.1088/0004-637X/709/2/1195}, \href
  {https://ui.adsabs.harvard.edu/abs/2010ApJ...709.1195T} {709, 1195}

\bibitem[\protect\citeauthoryear{{Van de Vyvere}, {Gomer}, {Sluse}, {Xu},
  {Birrer}, {Galan}  \& {Vernardos}}{{Van de Vyvere}
  et~al.}{2022}]{Vyvere_2022}
{Van de Vyvere} L.,  {Gomer} M.~R.,  {Sluse} D.,  {Xu} D.,  {Birrer} S.,
  {Galan} A.,   {Vernardos} G.,  2022, \mn@doi [\aap]
  {10.1051/0004-6361/202141551}, \href
  {https://ui.adsabs.harvard.edu/abs/2022A&A...659A.127V} {659, A127}

\bibitem[\protect\citeauthoryear{Varma, Fairbairn  \& Figueroa}{Varma
  et~al.}{2020}]{Dark_matter_CNN_1}
Varma S.,  Fairbairn M.,   Figueroa J.,  2020, Dark Matter Subhalos, Strong
  Lensing and Machine Learning (\mn@eprint {arXiv} {2005.05353})

\bibitem[\protect\citeauthoryear{Vegetti \& Koopmans}{Vegetti \&
  Koopmans}{2009}]{Vegetti2009}
Vegetti S.,  Koopmans L. V.~E.,  2009, Monthly Notices of the Royal
  Astronomical Society, 392, 945

\bibitem[\protect\citeauthoryear{Vegetti, Koopmans, Bolton, Treu  \&
  Gavazzi}{Vegetti et~al.}{2010}]{Vegetti_2010}
Vegetti S.,  Koopmans L. V.~E.,  Bolton A.,  Treu T.,   Gavazzi R.,  2010,
  \mn@doi [Monthly Notices of the Royal Astronomical Society]
  {10.1111/j.1365-2966.2010.16865.x}, 408, 1969

\bibitem[\protect\citeauthoryear{Vernardos}{Vernardos}{2018}]{Sersic_size}
Vernardos G.,  2018, \mn@doi [Monthly Notices of the Royal Astronomical
  Society] {10.1093/mnras/sty3486}, 483, 5583–5594

\bibitem[\protect\citeauthoryear{Vernardos}{Vernardos}{2021}]{MOLET}
Vernardos G.,  2021, Simulating time-varying strong lenses (\mn@eprint {arXiv}
  {2106.04344})

\bibitem[\protect\citeauthoryear{Vernardos \& Koopmans}{Vernardos \&
  Koopmans}{2022}]{Vernardos2022}
Vernardos G.,  Koopmans L. V.~E.,  2022, \mn@doi [Monthly Notices of the Royal
  Astronomical Society] {10.1093/mnras/stac1924}, 516, 1347

\bibitem[\protect\citeauthoryear{Vernardos, Tsagkatakis  \& Pantazis}{Vernardos
  et~al.}{2020}]{Vernardos_2020}
Vernardos G.,  Tsagkatakis G.,   Pantazis Y.,  2020, \mn@doi [Monthly Notices
  of the Royal Astronomical Society] {10.1093/mnras/staa3201}, 499, 5641–5652

\bibitem[\protect\citeauthoryear{Virtanen et~al.,}{Virtanen
  et~al.}{2020}]{Scipy}
Virtanen P.,  et~al., 2020, \mn@doi [Nature Methods]
  {10.1038/s41592-019-0686-2}, \href {https://rdcu.be/b08Wh} {17, 261}

\bibitem[\protect\citeauthoryear{{Vogelsberger} et~al.,}{{Vogelsberger}
  et~al.}{2014}]{Cosmology4}
{Vogelsberger} M.,  et~al., 2014, \mn@doi [\nat] {10.1038/nature13316}, \href
  {https://ui.adsabs.harvard.edu/abs/2014Natur.509..177V} {509, 177}

\bibitem[\protect\citeauthoryear{{Wagner-Carena}, {Park}, {Birrer}, {Marshall},
  {Roodman}, {Wechsler}  \& {LSST Dark Energy Science
  Collaboration}}{{Wagner-Carena} et~al.}{2021}]{CNN_Hierarchical_Inference}
{Wagner-Carena} S.,  {Park} J.~W.,  {Birrer} S.,  {Marshall} P.~J.,  {Roodman}
  A.,  {Wechsler} R.~H.,   {LSST Dark Energy Science Collaboration} 2021,
  \mn@doi [\apj] {10.3847/1538-4357/abdf59}, \href
  {https://ui.adsabs.harvard.edu/abs/2021ApJ...909..187W} {909, 187}

\bibitem[\protect\citeauthoryear{Wagner-Carena, Aalbers, Birrer, Nadler,
  Darragh-Ford, Marshall  \& Wechsler}{Wagner-Carena
  et~al.}{2022}]{CNN_Substructure_Mass_Function}
Wagner-Carena S.,  Aalbers J.,  Birrer S.,  Nadler E.~O.,  Darragh-Ford E.,
  Marshall P.~J.,   Wechsler R.~H.,  2022, From Images to Dark Matter:
  End-To-End Inference of Substructure From Hundreds of Strong Gravitational
  Lenses (\mn@eprint {arXiv} {2203.00690})

\bibitem[\protect\citeauthoryear{Warren \& Dye}{Warren \&
  Dye}{2003}]{Semilinear_inversion}
Warren S.~J.,  Dye S.,  2003, \mn@doi [The Astrophysical Journal]
  {10.1086/375132}, 590, 673

\bibitem[\protect\citeauthoryear{Wen, Vicol, Ba, Tran  \& Grosse}{Wen
  et~al.}{2018}]{Flipout}
Wen Y.,  Vicol P.,  Ba J.,  Tran D.,   Grosse R.,  2018, Flipout: Efficient
  Pseudo-Independent Weight Perturbations on Mini-Batches,
  \mn@doi{10.48550/ARXIV.1803.04386}, \url {https://arxiv.org/abs/1803.04386}

\bibitem[\protect\citeauthoryear{{Wilde}, {Serjeant}, {Bromley}, {Dickinson},
  {Koopmans}  \& {Metcalf}}{{Wilde} et~al.}{2022}]{CNN_lens_detection}
{Wilde} J.,  {Serjeant} S.,  {Bromley} J.~M.,  {Dickinson} H.,  {Koopmans} L.
  V.~E.,   {Metcalf} R.~B.,  2022, arXiv e-prints, \href
  {https://ui.adsabs.harvard.edu/abs/2022arXiv220212776W} {p. arXiv:2202.12776}

\bibitem[\protect\citeauthoryear{{de Vaucouleurs}}{{de
  Vaucouleurs}}{1948}]{Vaucouleurs_paper}
{de Vaucouleurs} G.,  1948, Annales d'Astrophysique, \href
  {https://ui.adsabs.harvard.edu/abs/1948AnAp...11..247D} {11, 247}

\makeatother
\end{thebibliography}

\appendix

\section{Sample of simulated strongly lensed images} \label{Appendix_lens_images}

\begin{figure*}
\centering
\includegraphics[width=0.97\textwidth]{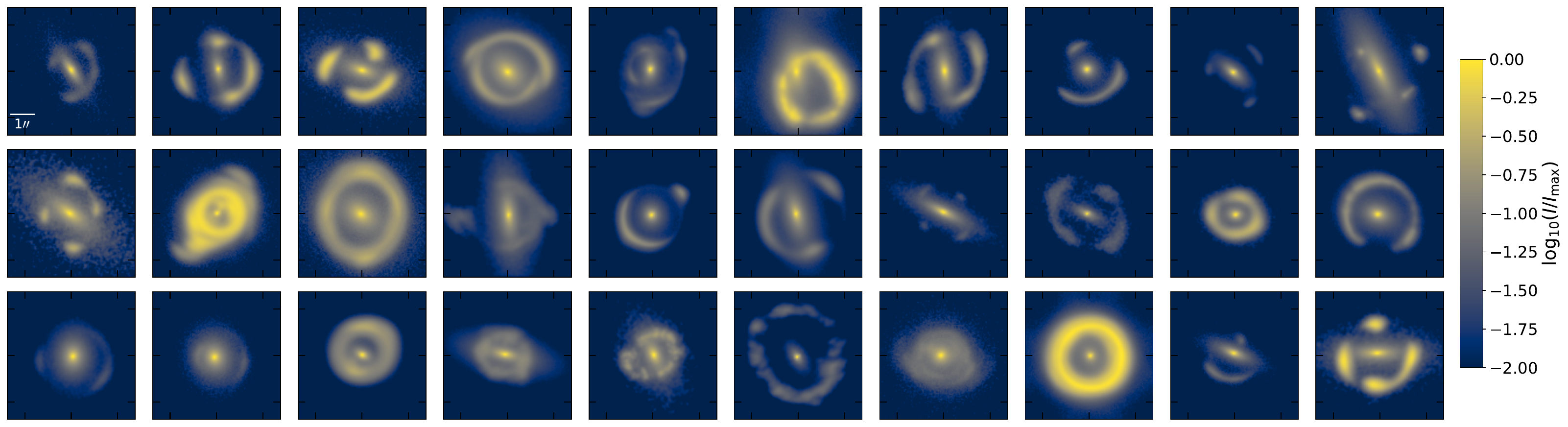}
\caption{Random sample of 30 mock images from our combined data set. Each image is scaled by its maximum brightness to better distinguish the arcs and rings of the lensed source from the lens light. All panels have the same dimensions as the SLACS data shown in Fig.~\ref{real_data}, i.e. $5.5''$ on a side with a resolution of $0.05''$ and $1''$ shown in the top leftmost panel.}
\label{lens_images}
\end{figure*}

Figure~\ref{lens_images} shows a sample of 30 mock strongly lensed images from the combined data set. Some of these lensed images appear grainy or offset due to strong perturbations to the smooth lens model. 

\section{Additional figures from different data sets} \label{appendix_extra_metrics}

\begin{figure*}
\centering
\includegraphics[width=0.95\textwidth]{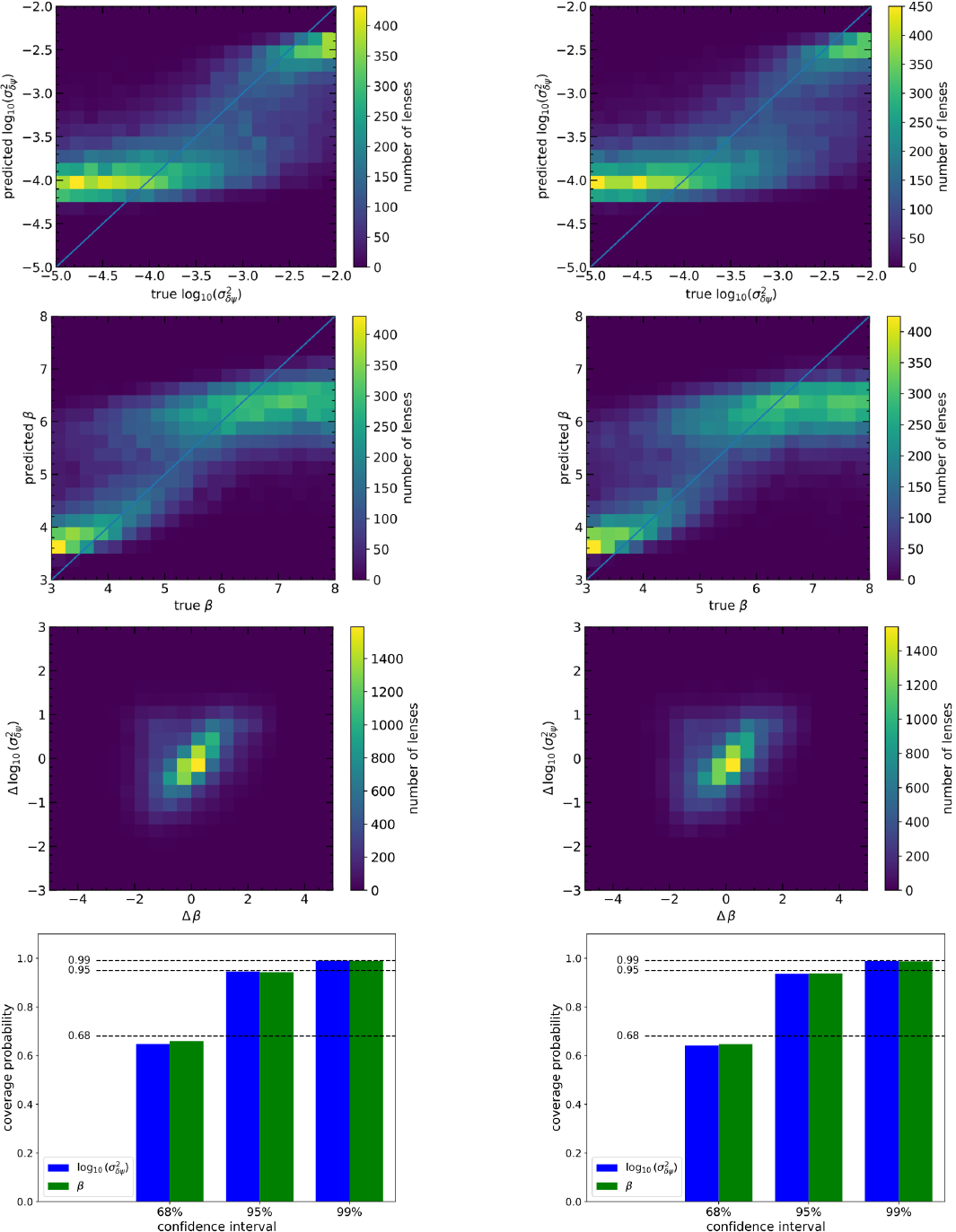}
\caption{Same as Fig.~\ref{loss_vs_epoch}, Fig.~\ref{confusion_matrices}, Fig.~\ref{difference_hist}, and Fig.~\ref{convergent_probability} from top to bottom, but for data set 1 (left) and data set 2 (right).}
\label{extra_metrics_fig}
\end{figure*}

\begin{figure*}
\centering
\includegraphics[width=0.97\textwidth]{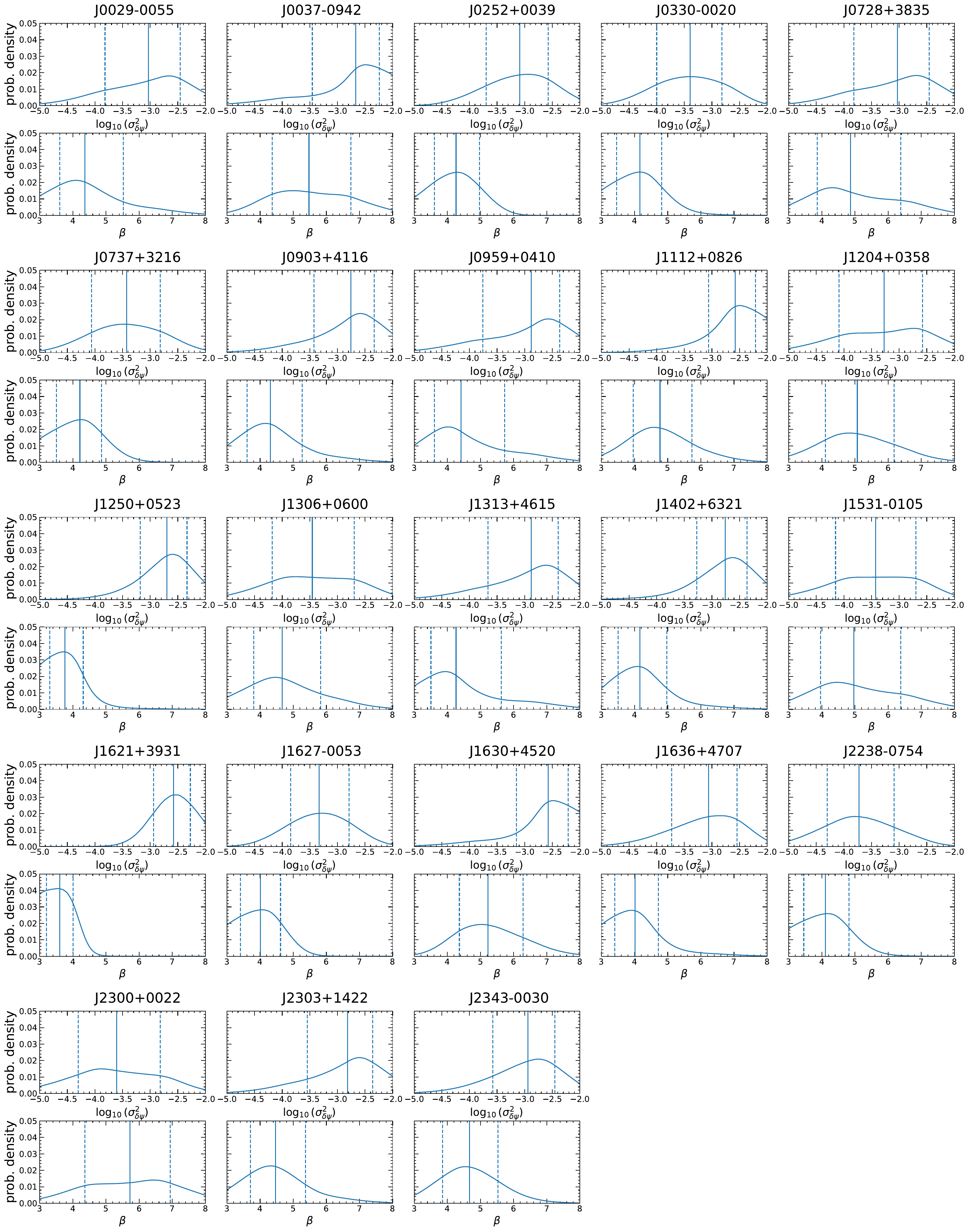}
\caption{Same as Fig.~\ref{probability_dist_of_data} but the average predicted probability distributions across the predictions of our three CNNs. The median values are shown by the vertical lines, and the 68 per cent confidence intervals are shown by the vertical dashed lines.}
\label{probability_dist_of_data_avg}
\end{figure*}

\begin{figure*}
\centering
\includegraphics[width=0.97\textwidth]{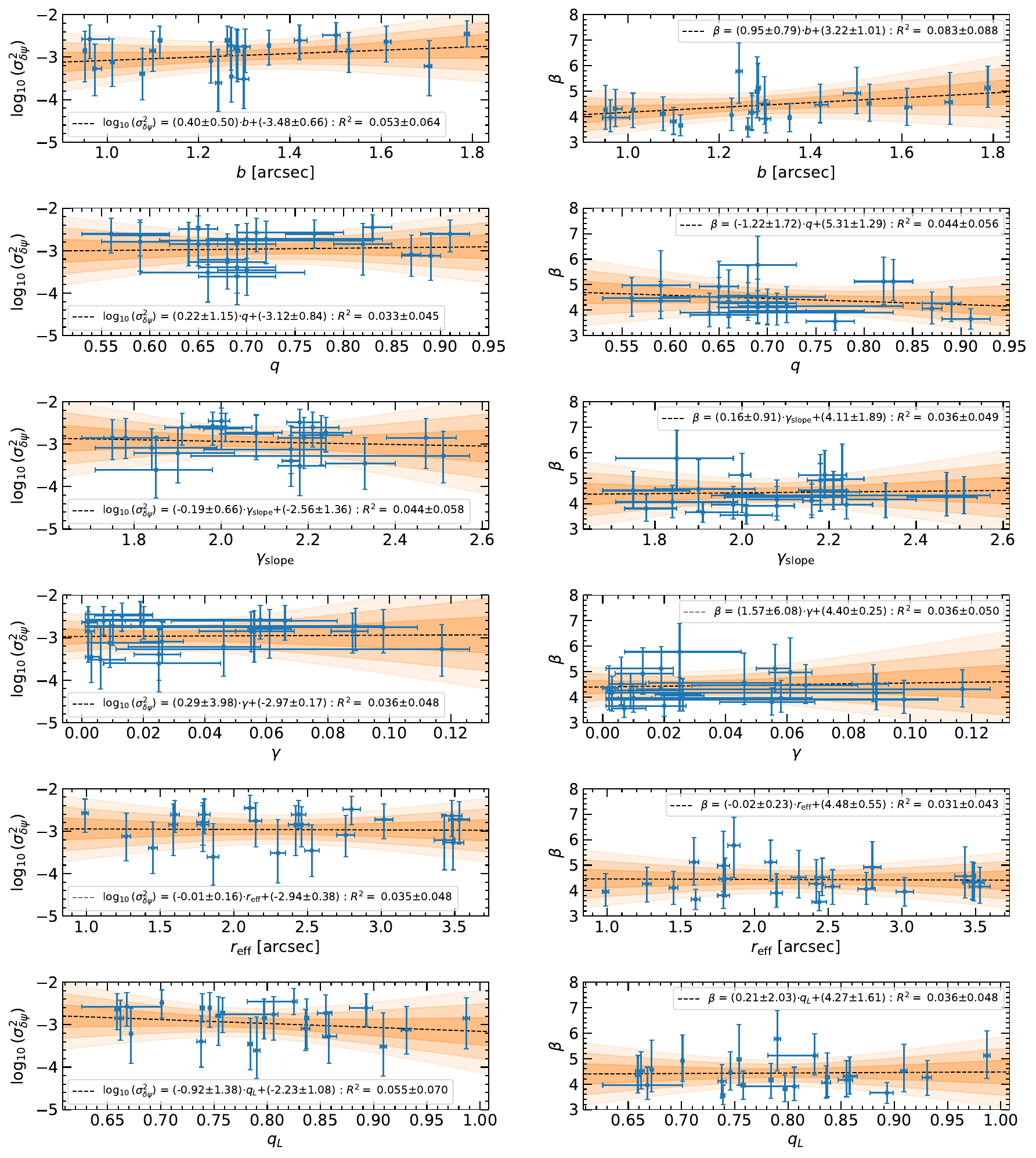}
\caption{Same as Fig.~\ref{A_and_beta_vs_lens_parameters_avg} but for data set 1}
\label{A_and_beta_vs_lens_parameters_data_set_1}
\end{figure*}

\begin{figure*}
\centering
\includegraphics[width=0.97\textwidth]{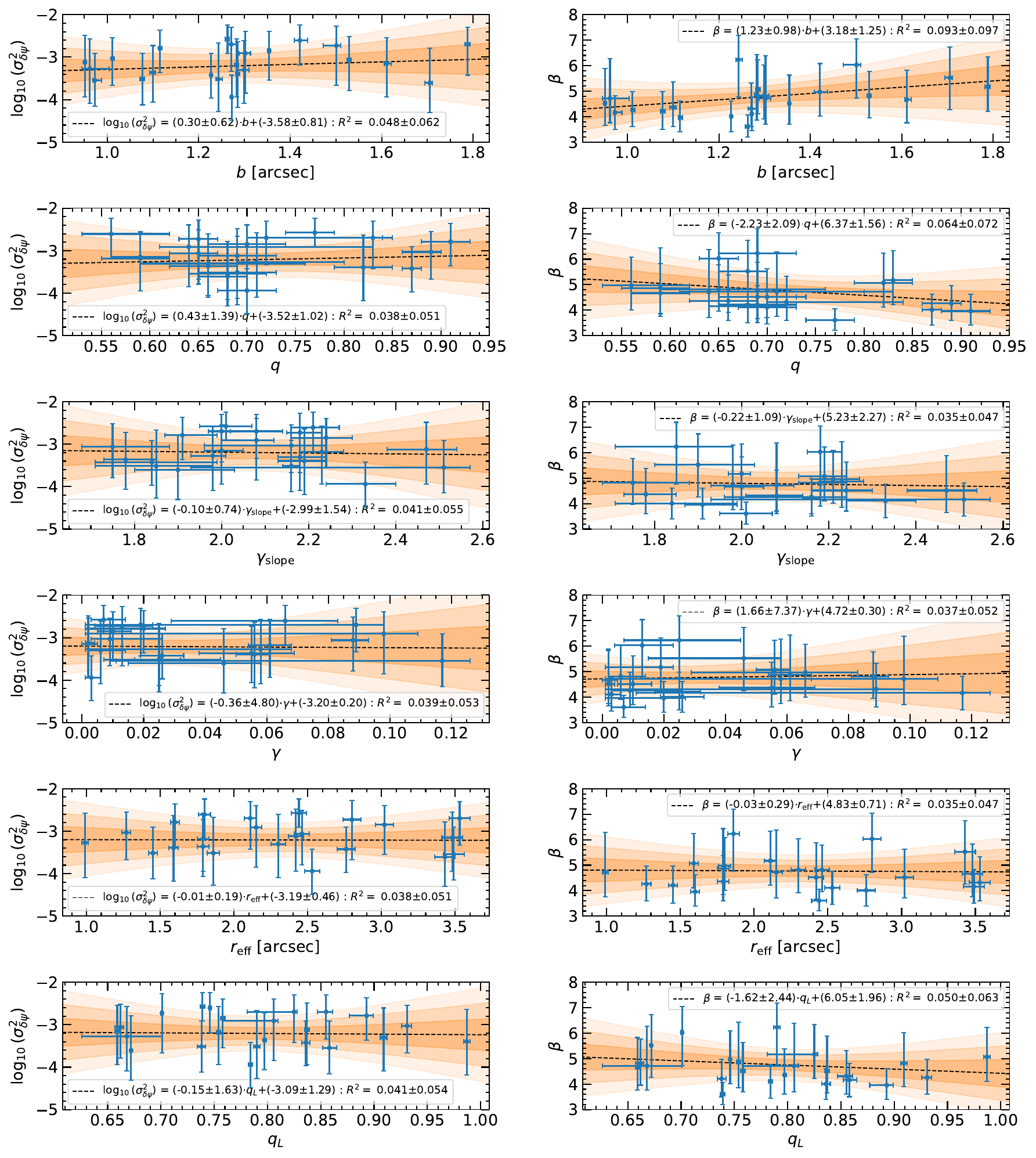}
\caption{Same as Fig.~\ref{A_and_beta_vs_lens_parameters_avg} but for data set 2}
\label{A_and_beta_vs_lens_parameters_data_set_2}
\end{figure*}

\begin{figure*}
\centering
\includegraphics[width=0.97\textwidth]{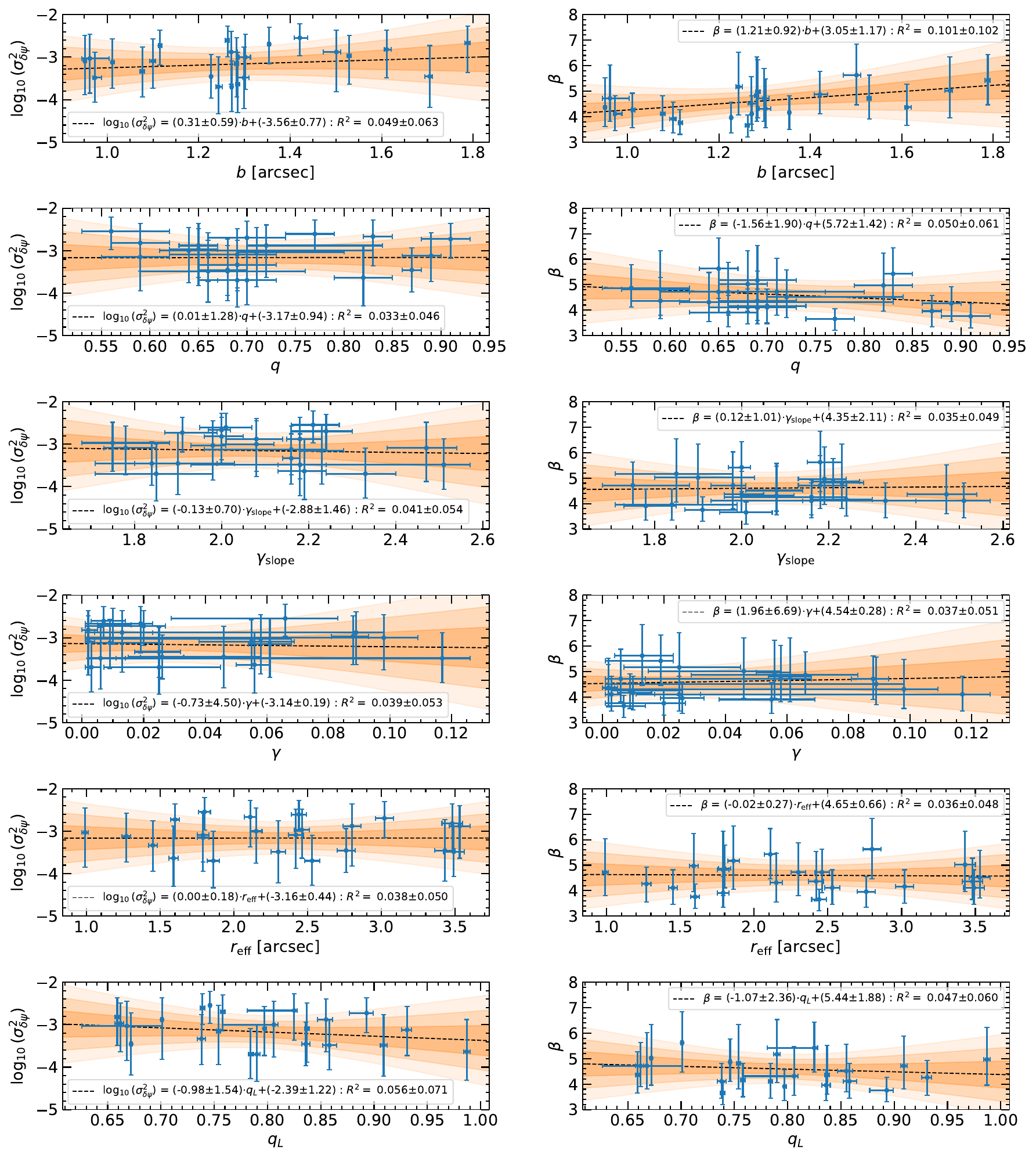}
\caption{Same as Fig.~\ref{A_and_beta_vs_lens_parameters_avg} but for the combined data set.}
\label{A_and_beta_vs_lens_parameters_data_set_combined}
\end{figure*}

Figure~\ref{extra_metrics_fig} shows additional figures from the CNNs trained on data set 1 and 2. This includes the MSE as a function of epoch, the mean predictions compared to the ground truth, difference histograms between the mean predictions and the ground truth, and the coverage probabilities for the 68, 95, and 99 per cent confidence intervals. Figure~\ref{probability_dist_of_data_avg} gives the average predicted probability distributions across our three CNNs for each of the SLACS lenses.
Figures~\ref{A_and_beta_vs_lens_parameters_data_set_1},~\ref{A_and_beta_vs_lens_parameters_data_set_2}, and~\ref{A_and_beta_vs_lens_parameters_data_set_combined} give our predictions compared to the smooth lensing parameters measured by~\citet{Shajib_2021} for our CNNs trained on data set 1, data set 2, and the combined data set respectively. Data set 1 (data set 2) has a root MSE of 20.1 (20.7) per cent for $\log_{10}(\sigma_{\delta\psi}^2)$ and 19.5 (20.1) per cent for $\beta$. The relative contribution of the epistemic uncertainties compared to the aleatoric uncertainty for data set 1 (data set 2) is 9.0 (8.9) per cent for $\log_{10}(\sigma_{\delta\psi}^2)$ and 5.5 (5.6) per cent for $\beta$.

\bsp	% typesetting comment
\label{lastpage}
\end{document}